\documentclass[aps,superscriptaddress,onecolumn,a4paper,preprint]{revtex4}
\usepackage[colorlinks=true, pdfstartview=FitV, linkcolor=blue, citecolor=red, urlcolor=black]{hyperref}
\usepackage{graphicx}
\usepackage[all]{xy}
\usepackage{amsmath}
\usepackage{amssymb}
\usepackage{tensor}
\newcommand{\be}{\begin{equation}}
\newcommand{\ee}{\end{equation}}
\newcommand{\ben}{\begin{eqnarray}}
\newcommand{\een}{\end{eqnarray}}
\newcommand{\bes}{\begin{subequations}}
\newcommand{\ees}{\end{subequations}}
\def\bal#1\eal{\begin{align}#1\end{align}}

\newcommand{\bfi}{\begin{figure}}
\newcommand{\efi}{\end{figure}}
\newcommand{\bc}{\begin{center}}
\newcommand{\ec}{\end{center}}

\newcommand{\arctanh}{\mbox{arctanh}}

\newcommand{\arcsech}{\mbox{arcsech}}
\newcommand{\sgn}{\mbox{sgn}}

\newcommand{\sech}{\mbox{sech}}

\newcommand{\LL}{{\cal L}}

\newcommand{\p}{\partial}

\begin{document}

\title{Maxwell-scalar system and scalar field with impurity: novel approach to find nontrivial solutions}

\author{I. Andrade}
\email{igor.as@ufma.br}
\affiliation{Departamento de Licenciaturas em Ci\^encias Naturais/Biologia, Universidade Federal do Maranh\~ao, 65400-000 Cod\'o, MA, Brazil}

\author{D. Bazeia}
\email{bazeia@fisica.ufpb.br}
\affiliation{Departamento de F\'\i sica, Universidade Federal da Para\'\i ba,\\58051-970 Jo\~ao Pessoa, PB, Brazil}

\author{M.A. Marques}
\email{marques@cbiotec.ufpb.br}
\affiliation{Departamento de Biotecnologia, Universidade Federal da Para\'\i ba, 58051-900 Jo\~ao Pessoa, PB, Brazil}

\author{R. Menezes}
\email{rmenezes@dcx.ufpb.br}
\affiliation{Departamento de Ci\^encias Exatas, Universidade Federal
da Para\'{\i}ba, 58297-000 Rio Tinto, PB, Brazil}

\begin{abstract}In this work, we investigate a Maxwell-scalar model that couples the scalar and gauge fields through the electric permittivity and another model, in which the scalar field lives in the presence of impurity. By considering a single spatial dimension, we determine the conditions under which the model with impurity can be seen as an effective model for the Maxwell-scalar system, having similar solutions. This correspondence shows that the impurity can be used to describe the presence of a charge density, and we use it to verify that the impurity-free case, which supports minimum energy configurations, is related to the case of point charges. We also investigate a class of impurities which modifies the core of the scalar field, and find the corresponding nontrivial charge densities and electric fields. In particular, the asymptotic behavior in terms of the impurity is also studied, leading to solutions exhibiting long-range or quasi-compact profile.
\end{abstract}

\maketitle

\section{Introduction}
The classical electromagnetism predicts that static charges have electric fields associated to them described by the so-called Gauss' law, $\nabla\!\cdot\! (\epsilon\,\textbf{E})=\varrho$, where $\varrho$ denotes the charge density and $\epsilon$ represents the permittivity of the system in the linear case \cite{jackson}. This equation implies that the electric field depends on the number of dimensions. For instance, it is well known that, in three spatial dimensions, the electric field generated by a static point charge is proportional to the inverse-square distance from the origin. However, in a single spatial dimension, its intensity is uniform, with a discontinuity at the charge location.

From the point of view of field theory, charged systems can be studied with the addition of the gauge field in the Lagrangian density via the Maxwell term \cite{landau}. Since in the present work we shall study the Maxwell term coupled to a scalar field in $(1,1)$ spacetime dimensions, it is of interest to recall that, in Refs. \cite{S1,S2}, Schwinger has studied the Maxwell system minimally coupled to fermions in $(1,1)$ spacetime dimensions, searching for exact solutions. The investigation obtained interesting results, unveiling that a vector gauge field can imply a nonzero mass particle, reinforcing the importance of gauge fields to describe short-range interactions. The Schwinger model has been further studied in several different directions, in particular, as a lattice model to describe classical-quantum computation using quantum computers \cite{QC}. More recently, the Maxwell term has been also used to include a non-minimal coupling of the gauge field with a real scalar field, to describe the behavior of electrically charged structures in systems described by point charges \cite{eletricloc,dipole,morris2}. As we shall show below, the Maxwell-scalar model is different from the Schwinger model, with the non-minimal coupling with a real scalar field modifying the medium where the gauge field evolves, bringing interesting new effects.

In Ref. \cite{eletricloc}, the Maxwell-scalar model was studied in the presence of a single charge, and in \cite{dipole} an electric dipole was considered. In the second case, after taking advantage of the bipolar geometry of the system, the investigation unveiled an interesting way to find stable minimum energy configurations connecting the two charges of opposite signs that form the dipole. In the present work, we want to go beyond the presence of point charges and investigate more general systems, described under the action of continuum charge distributions. This leads us to a much more intricate situation, and we shall implement the investigation following a novel procedure, in which we first describe solutions of a scalar field in the presence of impurities, and use them to map distinct electric fields and charge distributions in the Maxwell-scalar system. As one knows, scalar fields have been used over the years in the study of localized structures \cite{manton,vachaspati}. Among them, kinks are perhaps the simplest structures, and they arise under the action of a single real scalar field, as static solutions of the equation of motion that connect distinct minima of the potential. These objects are stable under small fluctuations and compatible with a first order framework based on energy minimization due to Bogomol'nyi \cite{bogo} and Prasad and Sommerfield \cite{PS}, sometimes referred to as BPS states. Kinks find several applications in Physics, such as in the study of braneworlds \cite{rs,dewolfe,csaki}, dilaton gravity \cite{zhong}, Bose-Einstein condensates \cite{bec1,bec2} and magnetic materials \cite{mag1,mag2}.

 The standard models in which localized structures appear usually do not account for spatial inhomogeneities that may be present due to impurities or external forces, being invariant under spatial translations. To get a more realistic scenario, which is, in general, more appropriate for practical applications, one may introduce impurities in the system. In the case of kink-like structures, many papers have dealt with this issue; see Refs.~\cite{imp1,imp2,imp3,imp4,imp5,imp6,imp7,imp8,imp9,imp10,imp11,imp12,imp13,imp14,imp15}. In particular, in Ref.~\cite{imp4}, it was shown that an attractive impurity may reflect the localized structure under specific conditions. More recently, in Refs.~\cite{imp10,imp11,imp12,imp13,imp14,imp15} the authors dealt with models where the impurities preserve BPS properties, with first order equations that simplify the problem. In this direction, in Ref. \cite{PRL} the study of scalar fields in the presence of BPS-preserving impurities has led to the appearance of spectral wall, where the scattering of kinks is importantly modified in a way similar to a hard-wall reflection. This phenomenon has been further considered in \cite{PRD,SW1,SW2} and in references therein.

Going to the direction of studying the Maxwell-scalar system, in this work we also investigate a model of a single real scalar field in the presence of impurities. By looking into its properties, such as the equation of motion, energy-momentum tensor and stability under small fluctuations, rescale of arguments and spatial translations, we find the conditions which allow the scalar field with impurities to be treated as an effective model for the Maxwell-scalar system. This correspondence between the Maxwell-scalar system and the scalar field model in the presence of impurity is the main result of this work, and it will be further explored below, to allow for the construction of nontrivial solutions for the Maxwell-scalar system. In this sense, since scalar fields are easier to be simulated on quantum computers \cite{QC2}, we think our results may stimulate new research concerning quantum computing \cite{QC,QC2}, to simulate the Maxwell-scalar system.

The search for a direct correspondence between the Maxwell-scalar system and the scalar field model was inspired by other possibilities, in particular, the well-know AdS/CFT correspondence, which relates Anti de Sitter spacetimes used in the description of quantum gravity in string theory, with quantum field theories formulated as Conformal Field Theory \cite{ads1,ads2,ads3,ads4}. This is also called a duality, because it engenders an
equivalence between two different descriptions of the physics in distinct dimensions, with the gravitational theory being defined in a higher dimensional spacetime containing at least the same dimensions of the particle theory, with the addition of one extra dimension of infinite extent. Another correspondence of interest, based on symmetry transformations identified in Ref. \cite{jac1}, was explored in \cite{B1} within the context of Galileo invariant system in $(d,1)$ spacetime dimensions and its direct connection with the dynamics of a relativistic d-brane in $(d+1,1)$ dimensions. A similar study was also reported in \cite{jac2}, in the case of fluid mechanical systems that enjoy a hidden higher-dimensional dynamical symmetry; see, e.g., Ref. \cite{jac} for more information on the subject. There are other kinds of correspondence in physics; a more recent one, now considering the same number of spacetime dimensions, was unveiled in \cite{olmo1}, through the identification of a novel procedure to map the field equations of some  nonlinear Ricci-based metric-affine theories of gravity, with the addition of scalar matter described by a given Lagrangian, into the field equations of general relativity coupled to another scalar field model. This correspondence has been used in several distinct ways, and it was recently considered to investigate models of spherical boson stars in Palatini $f(R)$ gravity and also, to study their collapse under perturbation \cite{olmo2}. Since it connects different theories with the same number of spacetime dimensions, it is in this sense similar to the correspondence to be explored in this work, which appears in the presence of a single spatial dimension.

In order to implement the present investigation, in Sec. \ref{mod} we introduce the models and expose the methodology used to uncover the main result of the present work. To do that, in Sec.~\ref{secmaxscalar} we first deal with the Maxwell-scalar system in $(1,1)$ dimensions, in which a scalar field is coupled to the gauge field via the electric permittivity. Then, in Sec.~\ref{secimpurities}, we study the scalar field model in the presence of impurity, which is controlled by a function depending explicitly on the spatial coordinate. There, we  determine the conditions under which this model can be seen as an effective theory for the Maxwell-scalar system, unveiling a direct correspondence between the two models. We then illustrate how this correspondence works investigating several distinct examples in Sec. \ref{results}. In particular, in Sec.~\ref{secpoint} we describe the case of a point charge and show that it is equivalent to the impurity-free model. Next, in Sec.~\ref{secimptocharged} we present a class of impurities that modifies the core of the solutions and find the corresponding charge distributions and electric fields in the Maxwell-scalar system. The study goes on in Sec.~\ref{secasympt}, where we investigate how the asymptotic behavior of the impurity modifies the tail of the scalar field and of the electric field that arises in the corresponding charged model. As we shall show, an interesting possibility is the construction of scalar field solution with power law tail, which can be used in collisions, to investigate how the power law tail may change the physics of entities that engender long-range interactions; see, e.g., Ref. \cite{Cold} for recent information on physical systems that engender long-range interactions. Our work ends in Sec.~\ref{secconclusions}, where we present some final remarks and discuss distinct perspectives of future investigations.

\section{Models and Main Result}\label{mod}

Let us start investigating the Maxwell-scalar system in Sec. \ref{secmaxscalar} and then the scalar field coupled to impurity in Sec. \ref{secimpurities}. Methodologically, we shall investigate the possibility of finding solutions in the presence of charge distributions in the Maxwell-scalar system, and of impurities in the scalar field model. In particular, we shall also deal with stability, because we want to find nontrivial solutions that are stable from the point of view of the Derrick theorem and the classical or linear stability as well. As it will be shown below, the study of stability requires the equations of motion and the energy-momentum tensors, and they will then be used to unveil the conditions for a direct connection between impurity and charge density, under which the correspondence between the two models can be established.

\subsection{Maxwell-scalar system}\label{secmaxscalar}

We first consider the Maxwell system in a medium with generalized electric permittivity in flat spacetime with $(1,1)$ spacetime dimensions, with $x^\mu=(x^0=t,x^1=x)$. The associated Lagrangian density is
\be\label{lel}
\LL_{el} = - \frac{\epsilon(\phi)}{4}F_{\mu\nu}F^{\mu\nu} +\frac12\partial_\mu\phi\partial^\mu\phi - A_\mu j^\mu.
\ee
Here we use standard notation, with the infinitesimal distance obeying $ds^2=\eta_{\mu\nu} dx^\mu dx^\nu$, where $\eta_{\mu\nu}$ is the metric tensor 
\be\eta_{\mu\nu}=\begin{pmatrix} 
	1 & 0 \\
	0 & -1  \\
	\end{pmatrix}.
\ee
This tensor and its inverse, $\eta^{\mu\nu},$ which has the same form above, can be used for raising and lowering Lorentz indices, such as $x_\mu=\eta_{\mu\nu} x^\nu$, $x^\mu=\eta^{\mu\nu}x_\nu$, $F_\mu^{\;\;\nu}=
\eta^{\nu\alpha}F_{\mu\alpha}$, etc.
Furthermore, $\phi=\phi(x,t)$ denotes the real scalar field, $A_\mu=(A_0(x,t),A_1(x,t))$ stands for the gauge field, $F_{\mu\nu}=\partial_\mu A_\nu - \partial_\nu A_\mu$ is the electromagnetic field strength tensor and $\partial_\mu=(\partial_0=\partial/\partial t, \partial_1=\partial/\partial x)$. Also, $\epsilon(\phi)$ represents the electric permittivity and $j^\mu=(j^0(x,t),j^1(x,t))$ is an external conserved current, obeying $\partial_\mu j^\mu=\partial j^0/\partial t+\partial j^1/\partial x=0$. We consider the light speed $c=1$ and use dimensionless coordinates and fields, for simplicity. Some issues related to the above Maxwell-scalar model were investigated previously in Refs.~\cite{eletricloc,dipole}, in the presence of one and two (dipole) point charges. Here, however, we want to go further and study the case of a continuous charge  distribution. Although this is much harder, in $(1,1)$ spacetime dimensions we have found an interesting possibility, directly connected with scalar field and impurities, which is fully explained below. Toward this goal, we write the equations of motion associated to \eqref{lel} as
\bes \label{eom}
\bal \label{scalareom}
\frac{\partial^2\phi}{\partial t^2}-\frac{\partial^2\phi}{\partial x^2}+\frac14\frac{d\epsilon}{d\phi}F_{\mu\nu}F^{\mu\nu} &= 0,\\ \label{maxwelleom}
\partial_\mu\left(\epsilon(\phi)F^{\mu\nu}\right) &= j^\nu.
\eal
\ees
We focus on the case with a single spatial dimension, for electrostatic configurations, with charge density $j^0=\varrho(x)$ and current $j^1=0$. The Gauss' law that arise from the $\nu=0$ component of \eqref{maxwelleom} is $\nabla\!\cdot\!(\epsilon \textbf{E})=\varrho$, in which the electric field is defined as $\textbf{E}=E\,\hat{x}$, where $\hat{x}$ denotes the cartesian unit vector and $E=F_{01} = -\partial A_0/\partial x$, for $A_1=0$. From this, we get
\be\label{eeps}
\textbf{E} = \frac{g(x)}{\epsilon(\phi)}{\rm sgn}(x)\,\hat{x},
\ee
with $g(x)={\rm sgn}(x)\int dx\,\varrho(x)$. The standard case, in which the scalar and gauge fields are decoupled, is recovered for $\epsilon=1$. In the general case, notice that the scalar field is present in the above expression via $\epsilon(\phi)$. One can use Eq.~\eqref{scalareom} combined with Eq.~\eqref{eeps} to get
\be\label{phieps}
\frac{\partial^2\phi}{\partial t^2}-\frac{\partial^2\phi}{\partial x^2}+ g^2(x)\,\frac{d}{d\phi}\left(\frac{1}{2\epsilon}\right)=0. 
\ee
Notice that this equation does not depend on the gauge field. Nevertheless, the signature of the electric charges still appears in the function $g(x)$.

Associated to the Lagrangian density \eqref{lel}, one can define the energy-momentum tensor
\be\label{tmunuel}
T_{\mu\nu} = \epsilon(\phi)\left(F_{\mu\lambda}\tensor{F}{^\lambda_\nu} +\frac14\eta_{\mu\nu}F_{\lambda\sigma}F^{\lambda\sigma}\right)+\p_\mu\phi\p_\nu\phi -\frac12\eta_{\mu\nu}\p_\lambda\phi\p^\lambda\phi.
\ee
Since the Lagrangian density \eqref{lel} is not invariant under spacetime translations due to the presence of the charge density $\varrho(x)$, the above expression is not conserved. Instead, it obeys 
\be\label{conserv}
\partial_\mu T^{\mu\nu} = j_\alpha F^{\alpha\nu}.
\ee
The components of Eq.~\eqref{tmunuel} are
\bes\label{tmunuelcomp}
\bal\label{surviving}
\rho_{el} &= T_{00} = \frac12\left(\frac{\partial\phi}{\partial t}\right)^2+\frac12\left(\frac{\partial\phi}{\partial x}\right)^2 +\frac{\epsilon(\phi)}{2}|\textbf{E}|^2,\\
\sigma_{el} &= T_{11} = \frac12\left(\frac{\partial\phi}{\partial t}\right)^2+\frac12\left(\frac{\partial\phi}{\partial x}\right)^2 -\frac{\epsilon(\phi)}{2}|\textbf{E}|^2,\\ \label{momdens}
p &= -T_{01} = -\frac{\partial\phi}{\partial t}\frac{\partial\phi}{\partial x},
\eal
\ees
where $\rho_{el}$, $\sigma_{el}$ and $p$ represent the energy density, stress and momentum density, respectively. The energy ${\cal E}_{el}$ can be obtained by integrating $\rho_{el}$ in the line. By using Eq.~\eqref{eeps}, one can rewrite the energy density and stress exclusively in terms of the scalar field and $g(x)$, as
\bes \label{T11T22}
\bal\label{rhoel}
\rho_{el} &= \frac12\left(\frac{\partial\phi}{\partial t}\right)^2+\frac12\left(\frac{\partial\phi}{\partial x}\right)^2 +\frac{g^2(x)}{2\epsilon(\phi)},\\
\sigma_{el} &= \frac12\left(\frac{\partial\phi}{\partial t}\right)^2+\frac12\left(\frac{\partial\phi}{\partial x}\right)^2 -\frac{g^2(x)}{2\epsilon(\phi)}.
\eal
\ees

We consider static solution, with $\phi=\phi(x)$. In this case, the equation of motion \eqref{phieps} changes to
\be\label{eomstatic}
\frac{d^2\phi}{dx^2} = g^2(x)\,\frac{d}{d\phi}\left(\frac{1}{2\epsilon}\right).
\ee
Moreover, one can see from Eq.~\eqref{momdens} that the momentum density vanishes, $p=0$, and that the energy density and the stress given in Eqs.~\eqref{T11T22} become
\bes 
\bal\label{rhoelstatic}
\rho_{el} &= \frac12\left(\frac{d\phi}{d x}\right)^2 +\frac{g^2(x)}{2\epsilon(\phi)},\\
\sigma_{el} &=\frac12\left(\frac{d\phi}{d x}\right)^2 -\frac{g^2(x)}{2\epsilon(\phi)}.
\eal
\ees
In this model, one cannot integrate the equation of motion \eqref{eomstatic} to get  first order equations. Usually, it can be obtained from the uniform stress condition. Nevertheless, from Eq.~\eqref{conserv}, one can show that $d\sigma_{el}/dx=-\varrho(x) E$, where $E$ is defined right above Eq.~\eqref{eeps}. Thus, for a general configuration of charges, we see that $\sigma_{el}$ depends explicitly on $x$, so the aforementioned condition for the stress is not ensured. Next, we study the stability of the static solutions under rescale of argument, spatial translations and small fluctuations.

\subsubsection{Stability}

We want to search for stable solutions, so it is important to first investigate how the Derrick's argument \cite{derrick} works in this model. To implement this possibility, we use the energy density \eqref{rhoelstatic} and make a rescale in the argument of the scalar field, $\phi(x)\to\phi^{\lambda}(x)=\phi(\lambda x)$. Considering $y=\lambda x$, the energy associated to the rescaled solution is
\be
\begin{aligned}
{\cal E}^\lambda_{el} &= \int_{-\infty}^{+\infty} dy\,\Bigg(\frac{\lambda}{2}\left(\frac{d\phi(y)}{dy}\right)^2 +\lambda^{-1}\frac{g^2(y/\lambda)}{2\epsilon(\phi(y))}\Bigg).
\end{aligned}
\ee
To minimize the above energy for $\lambda=1$, we first impose $d{\cal E}^\lambda_{el}/d\lambda\big|_{\lambda=1}=0$, which leads to
\bes\label{conditionel12}
\be\label{conditionel}
\int^{+\infty}_{-\infty} dy \left(\frac12\left(\frac{d\phi(y)}{dy}\right)^2 -\frac{1}{2\epsilon(\phi(y))}\frac{d}{dy}\left(yg^2(y)\right)\right) = 0.
\ee
This expression is a condition for $\lambda=1$ to be a critical point. To ensure that it is also a minimum, we further consider $d^2{\cal E}^\lambda_{el}/d\lambda^2\big|_{\lambda=1}>0$, which imposes
\be\label{conditionel2}
\int^{+\infty}_{-\infty} dy\, \frac{1}{\epsilon(\phi(y))} \frac{d^2}{dy^2}\left(y^2g^2(y)\right) > 0.
\ee
\ees
The conditions \eqref{conditionel12} ensure the stability under rescale, that is, under contractions and dilations. The explicit presence of the spatial coordinate in the aforementioned expressions may lead to change of sign in the integrands, so we cannot take the equality/inequality locally.

Since the model is not invariant under spatial translations, we can also extend Derrick's argument to investigate the stability under $x\to y=x+\lambda$ in the argument of the scalar field. In this case, we use Eq.~\eqref{rhoelstatic} to get
\be
{\cal E}^\lambda_{el} = \int^{+\infty}_{-\infty} dy \left(\frac12\left(\frac{d\phi(y)}{dy}\right)^2 +\frac{g^2(y-\lambda)}{2\epsilon(\phi(y))}\right).
\ee
By requiring $\lambda=0$ to be a minimum, we get the conditions
\bes\label{derricktranslel}
\bal
&\int^{+\infty}_{-\infty} dy \,\frac{1}{\epsilon(\phi(y))}\frac{d}{dy}\left(g^2(y)\right) = 0,\\
&\int^{+\infty}_{-\infty} dy \, \frac{1}{\epsilon(\phi(y))}\frac{d^2}{dy^2}\left(g^2(y)\right) > 0.
\eal
\ees
The above expressions ensure that the solution will not change its location, i.e., it will be stable against translations.

The conditions \eqref{conditionel12} and \eqref{derricktranslel} ensure that the solutions will always have the same form, avoiding contractions and dilations, and at the same position, without translating in the space. We can also investigate the linear stability of the system by taking small fluctuations around the static fields, in the form $\phi(x,t)=\phi(x)+\eta(x,t)$ and $A_\mu(x,t)=A_\mu(x)+\xi_\mu(x,t)$. We use $\tilde{F}_{\mu\nu}=\p_\mu \xi_\nu-\p_\nu \xi_\mu$ to denote the electromagnetic tensor associated to the fluctuations of the gauge field. By substituting them in \eqref{maxwelleom}, one obtains
\be
\p_\mu\left(\frac{d\epsilon}{d\phi}F^{\mu\nu}\eta +\epsilon(\phi)\tilde{F}^{\mu\nu}\right)= 0,
\ee
which admits the solution
\be\label{fluc}
\tilde{F}^{\mu\nu} = -\frac{d\ln(\epsilon)}{d\phi}F^{\mu\nu}\eta.
\ee
So, the fluctuations of the gauge field are directly related to the fluctuations of the scalar field. In this sense, we can use the fluctuations in the fields, the above expression and Eq.~\eqref{eeps} in Eq.~\eqref{scalareom} to obtain 
\be
\frac{\partial^2\eta}{\partial t^2}-\frac{\partial^2\eta}{\partial x^2}+g^2(x)\frac{d^2}{d\phi^2}\left(\frac{1}{2\epsilon}\right)\eta = 0.
\ee
We suppose that the scalar fluctuations have the form $\eta(x,t)=\sum_i\eta_i(x)\cos(\omega_i t)$ to get
\be\label{stabilityelectric}
-\frac{d^2\eta_i}{dx^2} +g^2(x)\frac{d^2}{d\phi^2}\left(\frac{1}{2\epsilon}\right)\eta_i = \omega_i^2\eta_i.
\ee
The system is stable if the eigenvalues are non-negative, $\omega_i^2\geq0$.

\subsection{Scalar field with impurity}\label{secimpurities}

The equations \eqref{lel}--\eqref{stabilityelectric} describe a system of electric configurations with charge density $\varrho(x)$ immersed in a medium with electric permittivity controlled by a scalar field. We remark that the equation of motion \eqref{phieps} has the form
\be\label{phiV}
\frac{\partial^2\phi}{\partial t^2}-\frac{\partial^2\phi}{\partial x^2}+ f(x)\frac{dV}{d\phi}=0.
\ee
It may appear from the following Lagrangian density, associated to a single real scalar field in the presence of impurity,
\be\label{lim}
\LL_{im} = \frac12\p_\mu\phi\p^\mu\phi - f(x)V(\phi).
\ee
In the above equation, $f(x)$ is a non-negative function that controls the impurity. The standard case has $\mathcal{L}= (1/2)\partial_\mu\phi\partial^\mu\phi - V(\phi)$, which is obtained for constant $f$, as one can redefine the potential to absorb the aforementioned constant. We can define the impurity as 
\be\label{impurityf}
{\cal I}(x) = f(x)-1.
\ee
So, the impurity-free scenario will be considered as the case in which ${\cal I}(x)=0$, which is attained for $f(x)=1$.

Interestingly, we see that $\LL_{im}$ supports the same scalar field solutions of $\LL_{el}$ in Eq.~\eqref{lel} if we take
\be\label{condeff}
f(x)=g^2(x)\quad\text{and}\quad V(\phi)=\frac{1}{2\epsilon(\phi)}.
\ee
Since the function $g(x)$ is defined as an integral of $\varrho(x)$ (see below Eq.~\eqref{eeps}) we notice that $f(x)$ simulates the effects of the charge configuration. In this sense, the inclusion of impurity in the form \eqref{impurityf} can be associated to the presence of electric charges in the Maxwell-scalar model in Eq.~\eqref{lel}. One can use ${\cal L}_{im}$ to introduce another energy-momentum tensor
\be\label{tmunuim}
\Theta_{\mu\nu} = \partial_\mu\phi\partial_\nu\phi - \eta_{\mu\nu}\left(\frac12\p_\mu\phi\p^\mu\phi - f(x)V(\phi)\right).
\ee
Notice that, due to the explicit presence of the spatial coordinate in the Lagrangian density, this energy-momentum tensor is not conserved; it obeys the equation
\be
\partial_\mu\Theta^{\mu\nu}=V\partial^\nu f.
\ee
From the $\Theta_{01}$ component of Eq.~\eqref{tmunuim}, we get the same momentum density as in Eq.~\eqref{momdens}; the other components are
\bes
\bal
\label{rhoim}
\rho_{im} &= \Theta_{00} = \frac12\left(\frac{\partial\phi}{\partial t}\right)^2+\frac12\left(\frac{\partial\phi}{\partial x}\right)^2 + f(x)V(\phi),\\ \label{sigmaim}
\sigma_{im} &= \Theta_{11} = \frac12\left(\frac{\partial\phi}{\partial t}\right)^2+\frac12\left(\frac{\partial\phi}{\partial x}\right)^2 - f(x)V(\phi).
\eal
\ees
The above components become the ones in Eq.~\eqref{T11T22} if the conditions \eqref{condeff} are satisfied. Thus, the scalar field model \eqref{lim} with impurities is an effective model for the electrically charged Maxwell system \eqref{lel}, as it presents the same solution and energy-momentum tensor. This identification allows that we use the effective model to calculate the scalar field profile, find the electric field via Eq.~\eqref{eeps} and calculate $A_0$ with ${\bf E}= -\nabla A_0$. Since $f(x)$ and $g(x)$ are related by Eq.~\eqref{condeff}, one can calculate the charge density associated to the function $f(x)$ that drives the impurity \eqref{impurityf} of the effective model through
\be\label{chargedens}
\varrho(x) = 2\delta(x)\sqrt{f(x)} + \frac{\sgn(x)}{2\sqrt{f(x)}}\frac{df}{dx}. 
\ee
Notice that $\delta(x)$ is present in the above expressions for impurities in which $f(0)\neq0$. Next, we study stability of the static solutions.

\subsubsection{Stability}

Since the model engenders impurity, let us investigate stability of the static solutions in this new scenario. First, we deal with the stability of the model \eqref{lim} under rescale. We proceed as in the previous model by taking $\phi(x)\to\phi^{\lambda}(x)=\phi(\lambda x)$. In this case, we get the conditions
\bes\label{conditionimp12}
\bal\label{conditionimp}
&\int^{+\infty}_{-\infty} dy \left(\frac12\left(\frac{d\phi(y)}{dy}\right)^2 -V(\phi(y))\frac{d}{dy}\left(yf(y)\right)\right) = 0,\\
\label{conditionimp2}
&\int^{+\infty}_{-\infty} dy\, V(\phi(y)) \frac{d^2}{dy^2}\left(y^2f(y)\right) > 0.
\eal
\ees
In the impurity-free scenario, the integrand of \eqref{conditionimp} is the stress for static fields (see Eq.~\eqref{sigmaim}). In this specific situation, stressless solutions satisfy the above conditions. However, in the presence of impurities, i.e., for ${\cal I}(x)\neq 0$ or $f(x)\neq1$, Eqs.~\eqref{conditionimp} and \eqref{conditionimp2} can be satisfied by solutions with non-vanishing stress.

As in the previous scenario, we also investigate the stability under spatial translations, $x\to y=x+\lambda$ in the argument of the scalar field. It leads us to the conditions
\bes\label{derricktranslim}
\bal
&\int^{+\infty}_{-\infty} dy \,V(\phi(y))\frac{df(y)}{dy} = 0,\\
&\int^{+\infty}_{-\infty} dy \, V(\phi(y))\frac{d^2f(y)}{dy^2} > 0.
\eal
\ees

Next, we study the linear stability by taking small fluctuations around the static solutions, $\phi(x)$, in the form $\phi(x,t)=\phi(x)+\sum_i\eta_i(x)\cos(\omega_it)$. By using this expression in Eq.~\eqref{phiV}, we obtain
\be\label{stabilityim}
-\frac{d^2\eta_i}{dx^2} +U(x)\eta_i = \omega_i^2\eta_i,
\ee
where the stability potential is
\be\label{stabpot}
U(x)=f(x) \frac{d^2V}{d\phi^2}.
\ee
The presence of $f(x)$ does not allow for the factorization of Eq.~\eqref{stabilityim} in terms of adjoint operators, so one cannot ensure stability in general. However, notice that the conditions \eqref{condeff} make the equations \eqref{conditionel12}, \eqref{derricktranslel} and \eqref{stabilityelectric} be exactly the same as \eqref{conditionimp12}, \eqref{derricktranslim} and \eqref{stabilityim}. So, the model with impurities \eqref{lim} engenders the very same stability of the one in Eq.~\eqref{lel}, except for the presence of the fluctuations of the gauge field, related to the scalar ones through Eq.~\eqref{fluc}.

As one can see from the above results, the scalar field model with impurity defined by Eq.~\eqref{lim} can be used to describe the solution for $\phi(x)$ and the corresponding energy density $\rho(x)$, and they can be considered together with the Eqs.~\eqref{eeps}, \eqref{condeff} and \eqref{chargedens} to calculate the electric field and charge density related to these solutions for the electrically charged model in Eq.~\eqref{lel}. This methodology, which is implemented under the process of searching for stable solutions, allows that we establish a direct correspondence between the Maxwell-scalar system and the scalar field model with impurity. Below we explore the correspondence studying several distinct examples, constructing the respective nontrivial solutions and investigating the associated asymptotic behavior.

Before doing that, however, one notices that in the light of the above correspondence, in the model with impurity we may make the scalar field complex to search for charged solutions of the Q-ball type, in a way similar to the study presented in Ref. \cite{QBB}. This includes extra degree of freedom, and suggests studying how the correspondence changes in this new situation. Conversely, in the Maxwell-scalar system investigated above, we can also think of changing the external charge density for a dynamical complex scalar field to be minimally coupled with the gauge field. This includes extra degrees of freedom and may lead us to the case of gauged Q-balls; see, e.g, Refs. \cite{QB,QB2} and references therein. In this case, an interesting issue concerns the study of how to change the scalar field model with impurity to keep the correspondence between the two models in the enlarged scenario.   

\section{Specific Results}\label{results}

In this section, we further consider the methodology developed in the above Sec. \ref{mod} to explore distinct possibilities to illustrate the correspondence between the two systems. We first deal with the impurity-free case and then the presence of impurities and their connections with charged configurations in the Maxwell-scalar system. We also study the behavior of the solutions, examining how the impurities can be used to control the asymptotic profile of the scalar field and the corresponding electric field.

\subsection{Point charges: the impurity-free case}\label{secpoint}
We start with the simplest case, in which the impurity is absent, with ${\cal I}(x)=0$ or $f(x)=1$, and with the scalar field static. The corresponding charge density obtained from Eq.~\eqref{chargedens} is $\varrho(x) = 2\delta(x)$, which stands for a point charge fixed at $x=0$. In this situation, we get the electric field ${\bf E} = \text{sgn}(x)\hat{x}/\epsilon(\phi)$. Notice that, in the standard case ($\epsilon=1$), the electric field is uniform (except for a sign) and engenders a discontinuity at $x=0$. As we shall see, the scalar field can be used to modify this behavior. To find the profile of $\phi(x)$, we take advantage of the effective model and consider the energy density in Eq.~\eqref{rhoim}, which takes the form $\rho = \left(d\phi/dx\right)^2\!/2 + V(\phi)$. It is well known from the BPS procedure \cite{bogo,PS} that one can introduce an auxiliary function $W=W(\phi)$ to show that, if the potential is $V(\phi) = \left(dW/d\phi\right)^2/2$, the energy is minimized to ${\cal E}=|W(\phi(+\infty))-W(\phi(-\infty))|$ for solutions obeying the first order equations
\be\label{fo}
\frac{d\phi}{dx} =\pm W_\phi.
\ee
The upper and lower signs can be related by the change $x\to-x$. In this specific situation, one can also show that Eq.~\eqref{conditionimp} leads to the stressless condition, $\sigma=0$, so the solution is stable under rescale. On the other hand, the model is invariant under spatial translations, $x\to x+\lambda$. Thus, Eqs.~\eqref{derricktranslim} do not make sense here. Regarding the linear stability, the presence of the function $W$ allows us to write the stability equation \eqref{stabilityim} in the form $S^\dagger S \eta_i = \omega_i^2 \eta_i$, in which $S=-d/dx + d^2W/d\phi^2$ and $S^\dagger=d/dx + d^2W/d\phi^2$. The presence of the adjoint operators $S$ and $S^\dagger$ ensures that the operator $S^\dagger S $ is non-negative, and this leads to the absence of negative eigenvalues, such that the ground state is the zero mode, with $\omega_0^2=0$ and $\eta_0={\cal N}\exp(\int dx\, d^2W/d\phi^2)$, where $\cal N$ is normalization constant, ensuring the linear stability of the model; see, e.g., Refs. \cite{DB,stabkink}.

To illustrate this case, we take $W(\phi) = \phi-\phi^3/3$, whose associated potential is
\be\label{phi4}
V(\phi) = \frac12\left(1-\phi^2\right)^2.
\ee
This potential engenders a maximum at $\phi=0$ and minima at $\phi=\pm1$. Using the first order framework, one can show that the first order equation \eqref{fo} with positive sign leads to the following solution and energy density
\be
\phi(x) = \tanh(x) \quad\text{and}\quad \rho(x) = \sech^4(x).
\ee
By integrating the energy density, one gets the energy ${\cal E}=4/3$, as expected. Since the scalar field profile is known explicitly, the electric field \eqref{eeps} becomes
\be
{\bf E} = \sech^4(x)\,\text{sgn}(x)\, \hat{x}.
\ee
The above expression shows that the generalized permittivity acts to localize the electric field, which is uniform (except for a sign) in the standard case $\epsilon=1$. Since ${\bf E} = -\nabla A_0$, we get 
\be
A_0(x) = \frac23-\frac13\left(2+\sech^2(x)\right)\text{sgn}(x)\tanh(x).
\ee
In this expression, we have taken the condition $A_0(\pm\infty) =0$ to obtain the integration constant. The stability of the model is driven by the Schrödinger-like equation \eqref{stabilityim}, whose associated stability potential in \eqref{stabpot} is
\be\label{stabpotPT}
U(x) = 4 - 6\,\sech^2(x),
\ee
which is the so-called modified Pöschl-Teller potential. Its bound states are $\omega_0^2=0$ and $\omega_1^2=3$, and the respective eigenfunctions are $\eta_0(x)=(\sqrt{3}/2)\,\sech^2(x)$ and $\eta_1(x)=(\sqrt{3/2})\,\sech(x)\tanh(x)$. It also supports another state, with $\omega_2^2 = 4$, such that $\eta_2(x) = 1-(3/2)\,\sech^2(x)$. Since there is no negative eigenvalue, the model is stable under small fluctuations.

In order to change the behavior of the electrically charged structure at the origin, we consider the model described by $W(\phi) = \phi^{2-1/p}/(2-1/p)-\phi^{2+1/p}/(2+1/p)$, whose associated potential presents two minima at $\phi_{\pm}=\pm1$ and another minimum at $\phi=0$. It has the form
\be\label{pmodel}
V(\phi) = \frac12\phi^2\left(\phi^{-1/p}-\phi^{1/p}\right)^2.
\ee
This model was proposed in Ref.~\cite{newglob}, and the parameter $p$ obeys $p=1,3,5,7,\ldots$, with $p=1$ leading us back to the potential \eqref{phi4}. The solution of Eq.~\eqref{fo} and its energy density are
\bal
&\phi(x) = \tanh^p\left(\frac{x}{p}\right),\\
&\rho(x) = \tanh^{2p-2}\left(\frac{x}{p}\right)\sech^4\left(\frac{x}{p}\right).
\eal
The energy depends on the parameter $p$, obeying the expression ${\cal E}=4p/(4p^2-1)$. The above scalar field solution can be substituted in Eq.~\eqref{eeps}, which leads to the following electric field
\be
{\bf E} = \tanh^{2p-2}\left(\frac{x}{p}\right)\sech^4\left(\frac{x}{p}\right)\,\text{sgn}(x)\, \hat{x}.
\ee
This expression shows that we have not only localized the electric field, making it vanish asymptotically, but also eliminated the discontinuity that is present at the origin in the standard case ($\epsilon=1$). The above expression leads to
\be
A_0(x) = \frac{2p}{4p^2-1} -\frac{p}{4p^2-1}\left(2 +(2p+1)\,\sech^2\left(\frac{x}{p}\right)\right)\text{sgn}(x)\tanh^{2p-1}\left(\frac{x}{p}\right),
\ee
which vanishes asymptotically.

\subsection{From impurities to charged configurations}\label{secimptocharged}

\begin{figure}[t!]
\centering
\includegraphics[width=7.5cm,clip]{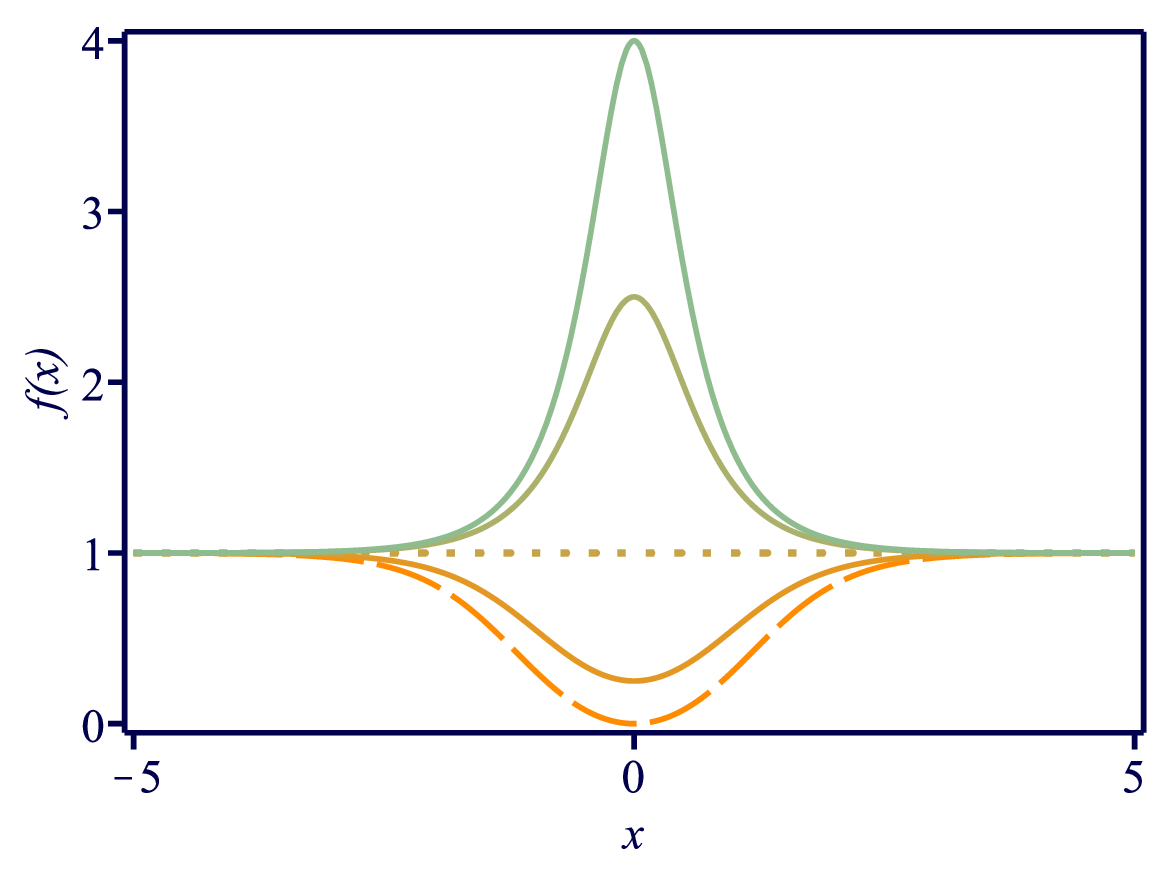}
\includegraphics[width=7.5cm,clip]{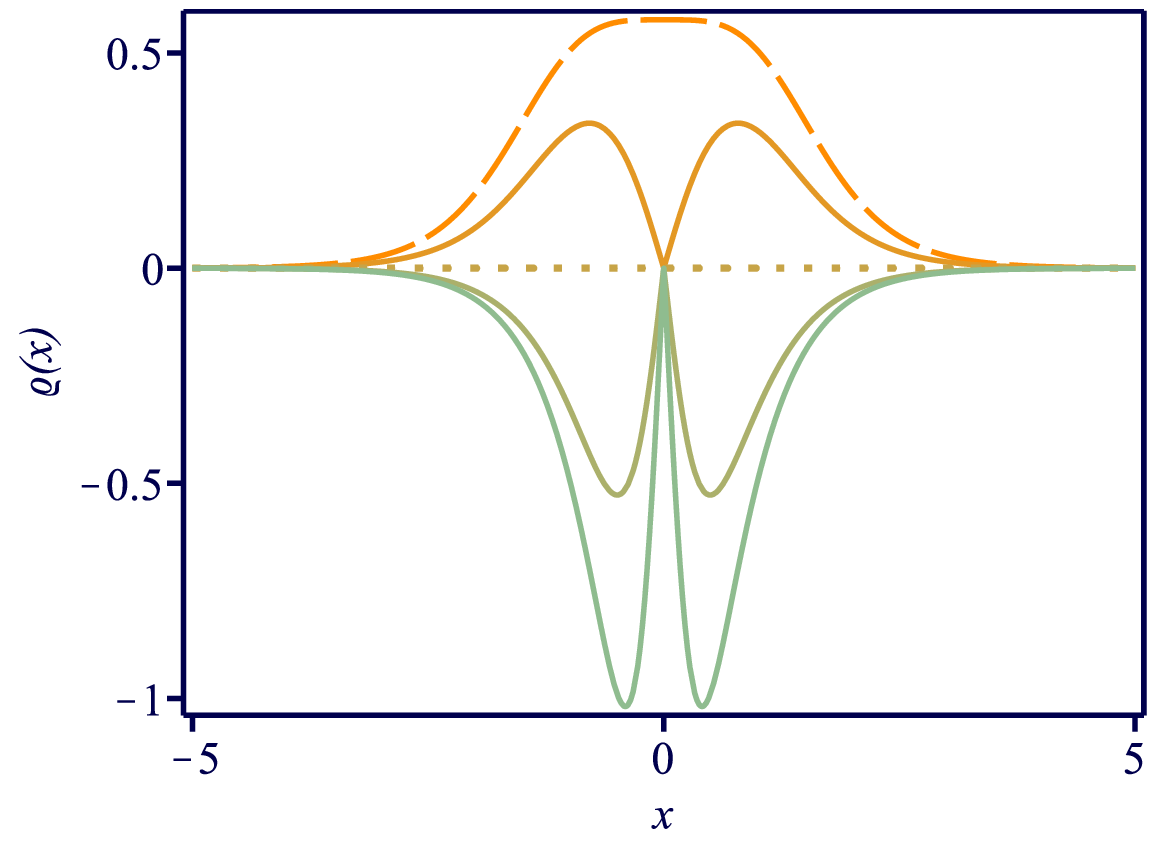}
\caption{The function $f(x)$ in Eq.~\eqref{impurity1} (left) and its corresponding charge density calculated from Eq.~\eqref{chargedens} (right) for $a=-2/3,-1/2,0,1$ and $2$. The dotted lines stand for $a=0$, the dashed ones represent the special case $a=-2/3$ and the solid ones are for the other values of $a$. The charge density engenders a $\delta(x)$ function for $a\neq-2/3$ that is not displayed here.}
\label{figimp1}
\end{figure}

Let us now illustrate the correspondence between impurities and charge densities. One can see from Eq.~\eqref{chargedens} that, for nonvanishing $f(0)$, there will always be a $\delta(x)$ in the charge density. We then introduce
\be\label{impurity1}
f(x) = \frac{2a+2+a\,\sech^2(x)}{2\left(a+1 -a\,\sech^2(x)\right)},
\ee
in which the parameter $a$ must obey $a\geq -2/3$, to ensure that $f(x)$ is non-negative. This function goes to unity asymptotically, i.e., $f(\pm\infty)=1$, for all $a$. The impurity-free model, with  ${\cal I}(x)=0$ or $f(x)=1$, is recovered for $a=0$. At $x=0$, $f(x)$ has a minimum for $a<0$ and a maximum for $a>0$, such that $f(0)=1+3a/2$. Notice that, for $a=-2/3$, we have $f(0)=0$, so $\delta(x)$ disappear in the charge density; see Eq.~\eqref{chargedens}. The impurity ${\cal I}(x)$ in Eq.~\eqref{impurityf} takes the form
\be\label{impa}
{\cal I}(x) = \frac{3a\,\sech^2(x)}{2\left(a+1 -a\,\sech^2(x)\right)}.
\ee
It vanishes asymptotically, modifying only the core of the structure. The function in Eq.~\eqref{impurity1} and its corresponding charge density from Eq. \eqref{chargedens} are displayed in Fig.~\ref{figimp1}. We remark that, for $a\neq-2/3$, the charge density presents a term with a contribution of $\delta(x)$, which is not depicted in the figure. 

We then investigate the model described by the $\phi^4$ potential in Eq.~\eqref{phi4}. We emphasize that, due to the presence of impurities, we cannot develop a first order framework in this case, thus we are now dealing with models the engender no BPS properties. In this sense, in the investigation to be considered below, the models are different from the ones investigated in Refs.~\cite{imp10,imp11,imp12,imp13,imp14,imp15}, which engender BPS properties. We are then entering a more difficult territory, which requires the use of the equation of motion \eqref{phiV}. For a static field, it reads
\be
\frac{d^2\phi}{dx^2} = -\frac{2a+2+a\,\sech^2(x)}{a+1 -a\,\sech^2(x)}\,\phi\left(1-\phi^2\right).
\ee
This equation supports the kink solution
\be\label{sol1}
\phi(x) = \frac{\sqrt{a+1}\tanh(x)}{\sqrt{1+a\tanh^2(x)}}.
\ee
\begin{figure}[t!]
\centering
\includegraphics[width=7.5cm,clip]{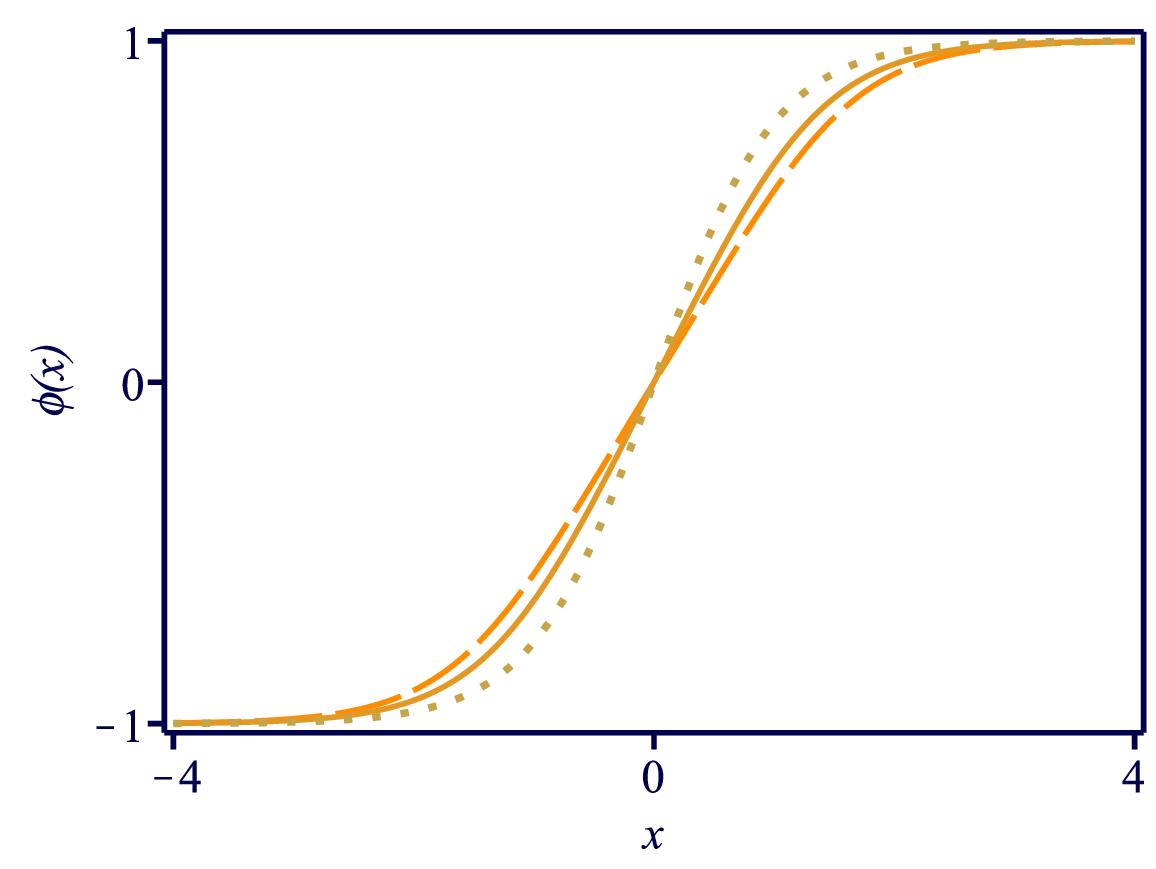}
\includegraphics[width=7.5cm,clip]{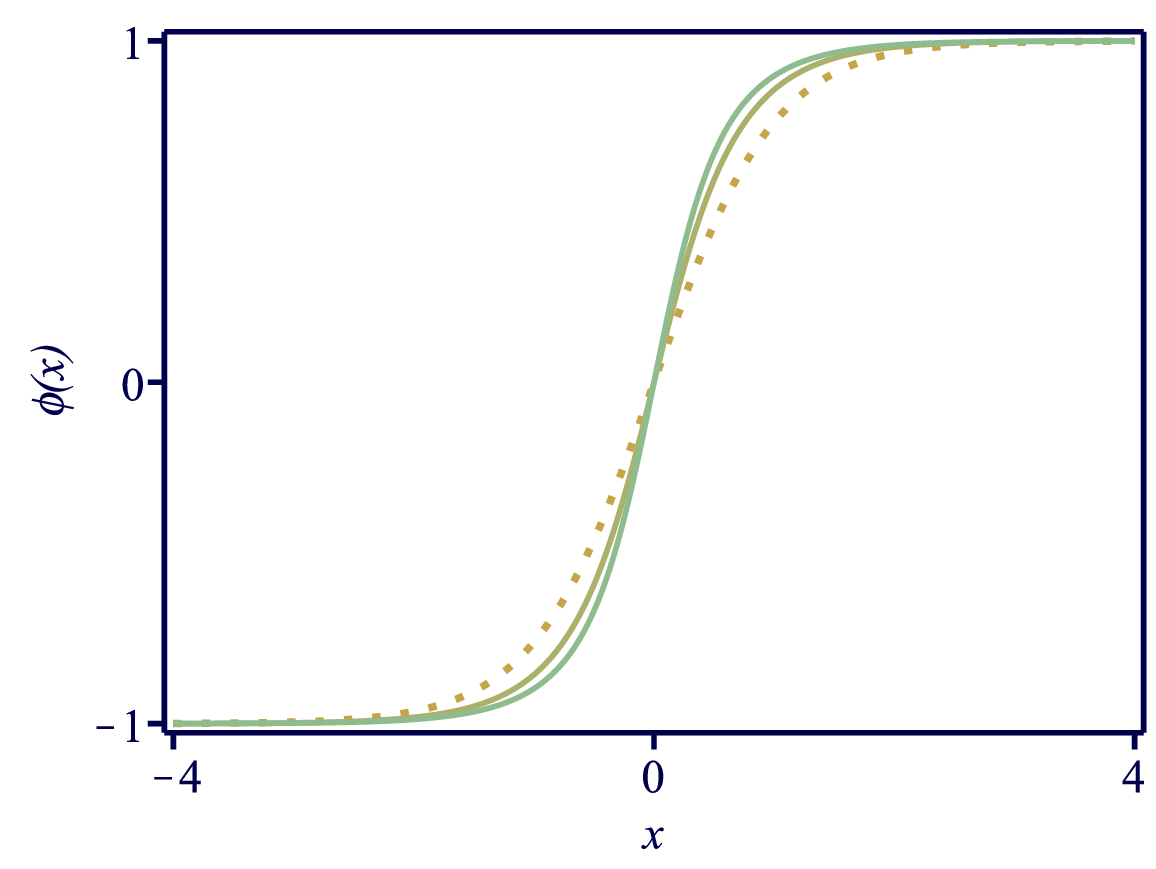}
\caption{In the left panel, one can see the solution $\phi(x)$ in Eq.~\eqref{sol1} for $a=-2/3,-1/2$ and $0$. In the right panel, it is displayed for $a=0$, $1$ and $2$. The dotted lines stands for $a=0$, the dashed one represent $a=-2/3$ and the solid ones are for the other values.}
\label{figsol1}
\end{figure}
Notice that, asymptotically, $\phi(\pm\infty) = \pm1$. So, even though the impurity is present, it does not modify the  boundary values of the kink, which connects the minima of the potential. In Fig.~\ref{figsol1}, we display the above solution for some values of $a$. The energy density can be calculated from Eq.~\eqref{rhoim}, which leads us to
\be\label{rho1}
\rho(x) = \frac{\left(4a+4+a\,\sech^2(x)\right)\sech^4(x)}{4\left(a+1-a\,\sech^2(x)\right)^3}.
\ee
\begin{figure}[t!]
\centering
\includegraphics[width=7.5cm,clip]{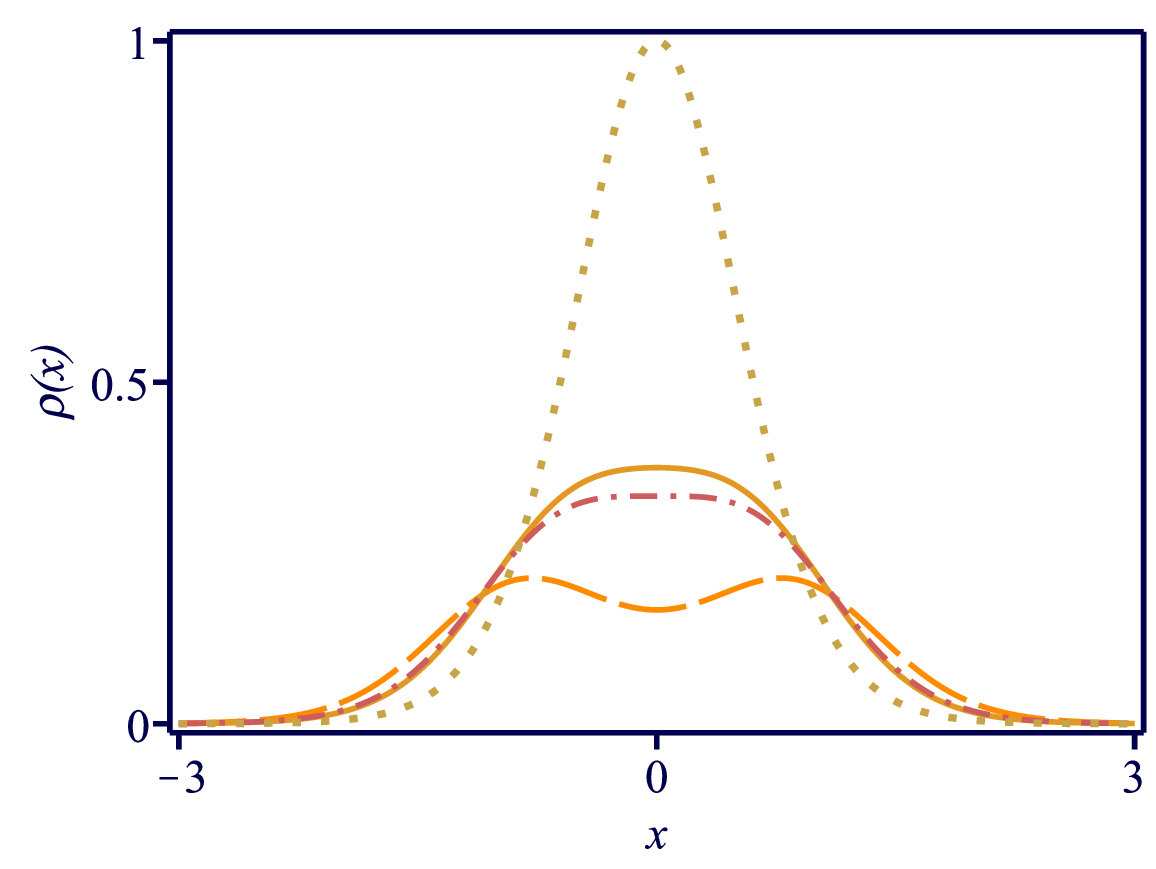}
\includegraphics[width=7.5cm,clip]{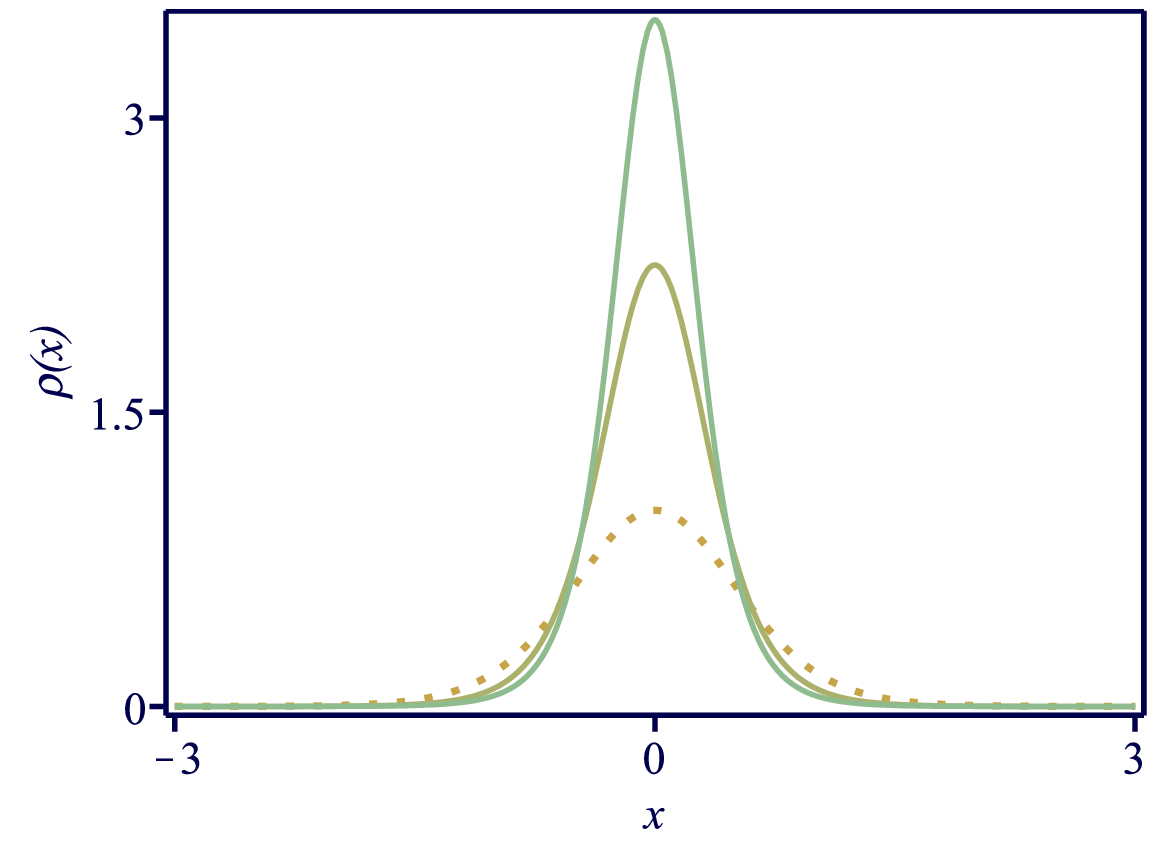}
\caption{The energy density $\rho(x)$ in Eq.~\eqref{rho1}. In the left panel, we depict it for $a=-2/3,-8/15,-1/2$ and $0$. In the right panel, we display it for $a=0,1$ and $2$. The lines follows the previous figures, except for the dash-dotted one, which stands for $a=-8/15$.}
\label{figrho1}
\end{figure}
This energy density may split for $a<-8/15$, with the two maxima located at $x=\pm\arcsech\big(\sqrt{-(8/7)(a+1)/a}\big)$. The impurity-free solution $\phi(x) = \tanh(x)$ and its corresponding energy density $\rho(x)=\sech^4(x)$ are recovered for $a=0$. In Fig.~\ref{figrho1} we display the above energy density for some values of $a$, including $a=-8/15$, which delimits the range in which the splitting occurs. By integrating the above expression, we get the energy
\be
{\cal E} = \frac{15a+1 +\left(8+3(a+1)(5a-3)\right)M(a)}{16a},
\ee
in which
\be
M(a) =
\begin{cases}
\displaystyle\arctanh\left(\sqrt{-a}\right)/\sqrt{-a}
,& a<0\\
\displaystyle\arctan\left(\sqrt{a}\right)/\sqrt{a}, & a>0.
\end{cases}
\ee
In the limit $a\to 0$, the energy is ${\cal E}=4/3$, as expected. As we have commented below Eq.~\eqref{conditionimp}, the stress does not vanish due to the presence of impurities. For the solution \eqref{sol1}, the stress \eqref{sigmaim} is
\be
\sigma(x) = -\frac{a\,\sech^6(x)}{4\left(a+1-a\,\sech^2(x)\right)^3},
\ee
which is negative for $a>0$, null for $a=0$ (impurity-free case) and positive for $a<0$.

\begin{figure}[t!]
\centering
\includegraphics[width=7.5cm,clip]{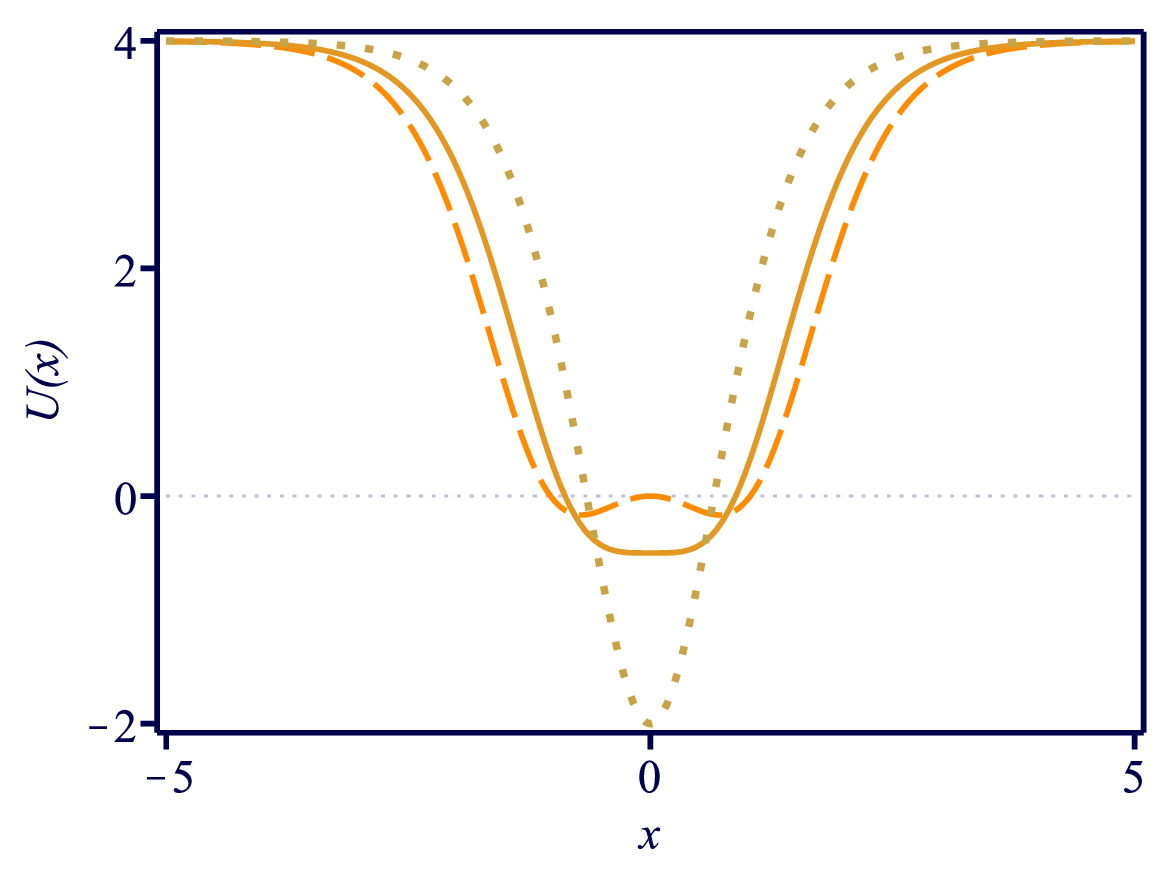}
\includegraphics[width=7.5cm,clip]{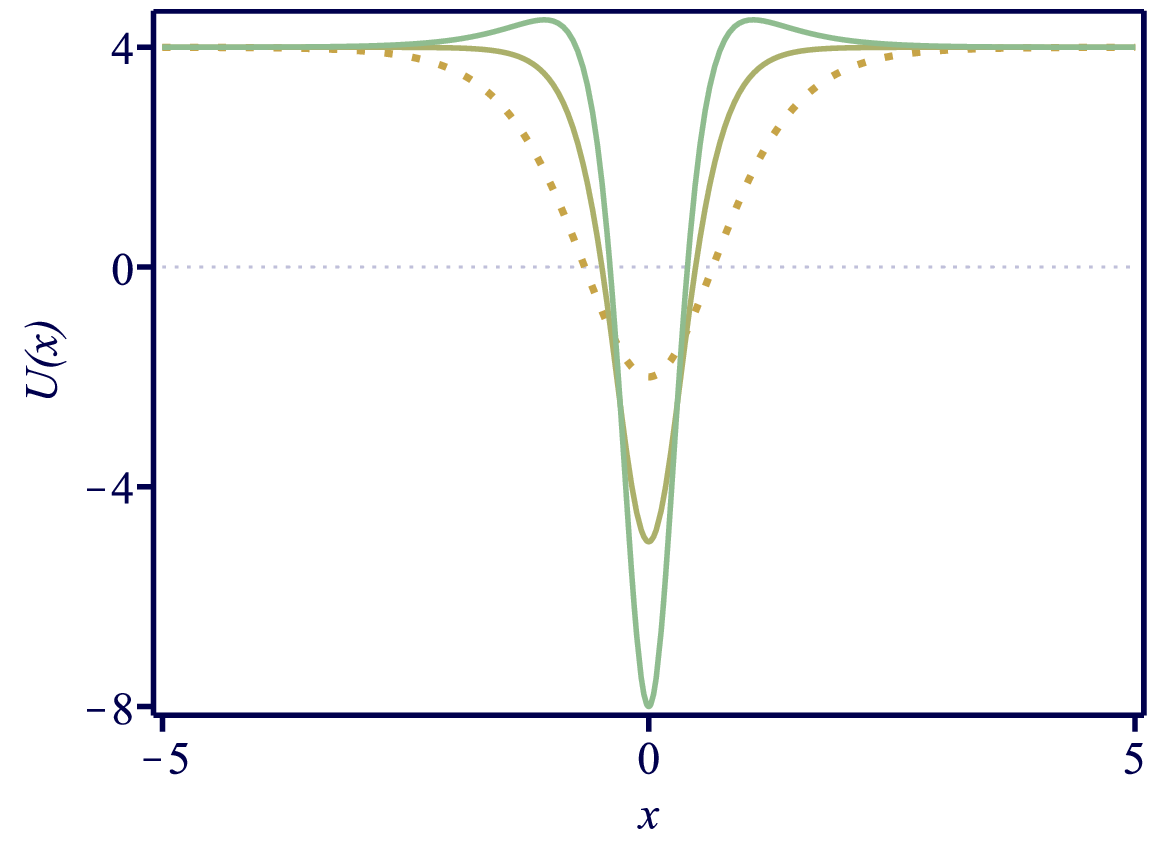}
\caption{The profile of the stability potential $U(x)$ in Eq.~\eqref{stabpot1}. In the left panel, we depict it for $a=-2/3,-1/2$ and $0$. In the right panel, we display it for $a=0$, $1$ and $2$. The lines follow the previous figures.}
\label{figstabpot1}
\end{figure}

\begin{figure}[t!]
\centering
\includegraphics[width=7.5cm,clip]{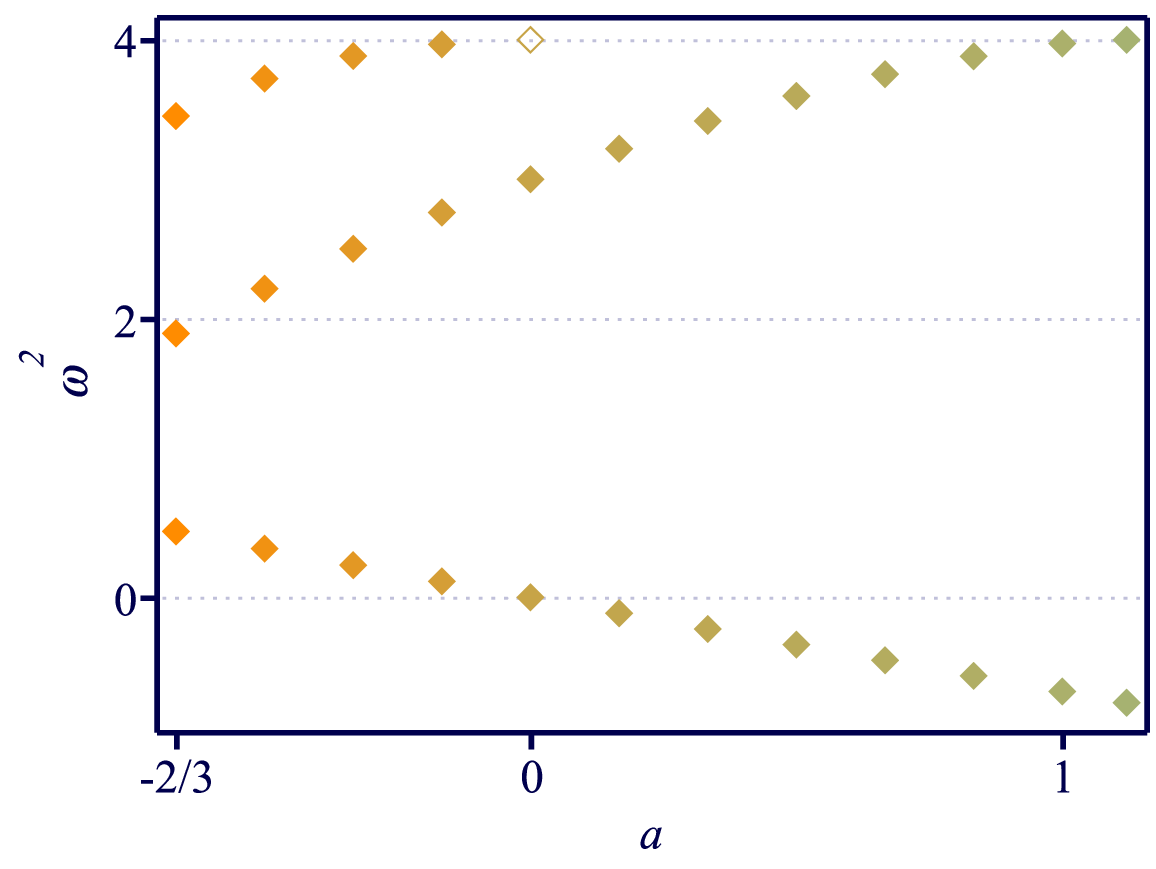}
\caption{The eigenvalues $\omega_i^2$ associated to the stability potential in Eq.~\eqref{stabpot1} for some values of $a$. The hollow diamond stands for $\omega_2^2=4$, which appears in the impurity-free case ($a=0$).}
\label{figbstates1}
\end{figure}

To investigate the stability, we take the potential in Eq.~\eqref{stabpot}, which leads us to
\be\label{stabpot1}
U(x) = \frac{4(a+1)^2 -2(a+1)(a+3)S^2 -a(2a+3)S^4}{\left(a+1-aS^2\right)^2},
\ee
where $S=\sech(x)$. In Fig.~\ref{figstabpot1}, we depict this potential for several values of $a$: in the left panel we display it for $a\leq0$, and in the right panel we do it for $a\geq0$. For $a=0$ we recover the modified P\"oschl-Teller potential in Eq.~\eqref{stabpotPT}. The point $x=0$ defines a global minimum (local maximum) for $a\geq-1/2$ ($a<-1/2$). The point $x=\pm\arcsech\big(\sqrt{(a^2-1)/(a(a+2))}\big)$ is a global minimum for $a<-1/2$ and a local maximum for $a>1$. To obtain the eigenvalues and eigenfunctions, one must solve Eq.~\eqref{stabilityim}. We use numeric procedures to calculate the bound states $\omega_i^2$ for some values of $a$, and display them in Fig.~\ref{figbstates1}. Notice that the zero mode does not appear for $a\neq0$ due to the absence of translational symmetry in the system. We see that, for $a<0$, this potential supports three bound states with positive eigenvalues. For $a>0$, the number of bound states changes with $a$: it has two bound states for $0<a\leq a_*$, with $a_*\approx 1.121$, and a single one for $a>a_*$. For both situations with positive $a$, negative eigenvalues always appear, giving rise to instability.

We have checked that the model is stable against contractions and dilations, obeying Eqs.~\eqref{conditionimp12}. The stability under translations depends on $a$. The case $a>0$ leads to instability, such that the kink is expelled from the region in which the impurity is present. On the other hand, for $a<0$, the conditions \eqref{derricktranslim} are attained, entrapping the solution at the center of the impurity. Interestingly, this is the same interval where the linear stability is ensured. Indeed, we can expand the stability potential \eqref{stabpot1} for $a\approx0$ and make the approximations $\eta_i=\eta_{i(0)}+a\,\eta_{i(1)}$ and $\omega_i^2=\omega_{i(0)}^2+a\,\omega_{i(1)}^{2}$, where the $\eta_{i(0)}$ and $\omega_{i(0)}^2$ are, respectively, the eigenfunctions and eigenvalues of modified Pöschl-Teller potencial \eqref{stabpotPT}, in the stability equation \eqref{stabilityim}. The procedure leads us to 
\be
\omega_{i(1)}^2 = \int^{+\infty}_{-\infty} dx\left(3S^2\left(4-5S^2\right)\right)\left|\eta_{i(0)}\right|^2.
\ee
So, we have $\omega_0^2=-24a/35$ and $\omega_1^2=3+48a/35$. This shows that the ground state $\omega_0^2$ is negative for $a>0$, which induces instability, and positive for $a<0$. Furthermore, in both cases the ground state is not the zero mode, as expected due to the explicit spatial dependence introduced by the impurity in the model.

The electric field in Eq.~\eqref{eeps} takes the form
\be
\textbf{E} = \sech^4(x)\sqrt{\frac{2a+2+a\,\sech^2(x)}{2\left(a+1 -a\,\sech^2(x)\right)^5}}\,\text{sgn}(x)\,\hat{x}.
\ee
From this expression, we get that $\textbf{E}$ is localized, vanishing asymptotically for any value of $a$. The behavior at the origin presents a discontinuity for $a>-2/3$. The case $a=-2/3$, however, leads to $\textbf{E}(0)=0$, eliminating the discontinuity and making the electric field smooth.

\subsection{Asymptotic behavior}\label{secasympt}

An interesting feature related to the impurity investigated in Eq.~\eqref{impa} with the potential \eqref{phi4} is that its associated solution \eqref{sol1} engenders exponential tails, in the form $1\mp\phi(x) \propto \exp(-2|x|)$ for $x\to\pm\infty$. The electric field behaves asymptotically as $\textbf{E} \propto \exp(-4|x|)\hat{x}$ and the gauge field as $A_0(x) \propto \exp(-4|x|)$. This exponential tail is a consequence of the asymptotic behavior of the function that drives the impurity, $f(x)$, which is constant for points far away from the center of the solution. 

In order to better understand how an arbitrary $f(x)$ modifies the tails of the solution, we take the approximation $\phi(x) = v_+ - \phi_{as}(x)$ in Eq.~\eqref{phiV} for the static case, where $v_+$ is the minimum of the potential connected by the solution at $x\to+\infty$. We also take the classical mass $m^2=d^2V/d\phi^2\big|_{\phi=v_+}$ to be finite. The behavior of the electric field can be obtained via Eq.~\eqref{eeps}. Since we are interested in even $f(x)$, we only study the $x\to+\infty$ case here, as the analysis for $x\to-\infty$ is similar.

We now focus on investigating the case in which $f(x) \approx k^2$ asymptotically, where $k$ is a positive constant. This leads to 
\be
\phi_{as}(x) \propto \exp(-k mx)\quad\text{and}\quad\textbf{E}\propto\exp(-2kmx)\, \hat{x}.
\ee
So, the exponential tails in the solution indeed arise for $f(x)$ with positive constant asymptotic behavior, matching with the result obtained for the solution \eqref{sol1}.

Another possibility is to consider $f(x) \approx k^2 x^n$ asymptotically, with $n\geq-2$ to ensure that the scalar field solutions connect the minima of the potential $V(\phi)$. In this situation, the expression describing the tail depends on $n$. For $n>-2$, we get
\bes
\bal
&\phi_{as}(x) \propto \frac{1}{x^{n/4}} \exp\left(-\frac{2km}{n+2}\,x^{1+n/2}\right),\\
&\textbf{E}\propto\exp\left(-\frac{4km}{n+2}\,x^{1+n/2}\right)\,\hat{x}.
\eal
\ees
The case $n=-2$ is special, leading to the power law behavior
\bes
\bal
&\phi_{as}(x) \propto \frac{1}{x^{(\sqrt{1+4k^2m^2}-1)/2}},\\
&\textbf{E}\propto \frac{1}{x^{\sqrt{1+4k^2m^2}}}\,\hat{x}.
\eal
\ees
These expressions lead to a slower (faster) falloff than the usual exponential one for negative (positive) $n$.

We can further suppose that $f(x)$ diverges exponentially, asymptotically, in the form $f(x) \approx k^2\exp(nx) $, with $n>0$. In this case, the scalar field solution and the electric field behave asymptotically as
\bes
\bal
&\phi_{as}(x)\propto\exp\left(-\frac{2km}{n}\,\exp\left(\frac{n}{2}\,x\right)\right),\\
&\textbf{E}\propto\exp\left(-\frac{4km}{n}\,\exp\left(\frac{n}{2}\,x\right)\right)\,\hat{x}.
\eal
\ees
We then see that they engender a double exponential falloff, which is faster than the usual exponential one.

As one can see from Eq.~\eqref{chargedens}, the charge density $\varrho(x)$ depends explicitly on the function $f(x)$. By integrating it over the real line, one gets the total charge, which depends only on the boundary values of $f(x)$. Since $f(x)$ is even, the charge is $Q=2f(+\infty)$. In this sense, depending on the asymptotic behaviors discussed above, the charge may vanish, or be finite or divergent.

\subsubsection{Long-range structures}\label{seclongrange}

We have shown that solutions with power law tails arise for impurities with inverse-quadratic falloff at infinity. As investigated in Ref.~\cite{highly}, this type of structure may engender long-range interactions. To illustrate this feature, one may take
\be\label{impuritylong}
f(x) = \frac{3}{2\left(x^2+1\right)},
\ee
which presents a maximum at $x=0$, with $f(0)=3/2$, and vanishes asymptotically in the form $f(x)\approx3/(2x^2)$. The static solution of Eq.~\eqref{phiV} which connects the minima of the potential \eqref{phi4} is
\be\label{sollong}
\phi(x) = \frac{x}{\sqrt{x^2+1}}.
\ee
So, it gets power law tails, as expected from the previous discussion. The energy density \eqref{rhoim} becomes
\be\label{rholong}
\rho(x) = \frac{5}{4\left(x^2+1\right)^{3}}
\ee
and the stress \eqref{sigmaim} reads $\sigma(x) = -(1/4)\left(x^2+1\right)^{-3}$. The energy is ${\cal E}=15\pi/32$.

\begin{figure}[t!]
\centering
\includegraphics[width=7.5cm,clip]{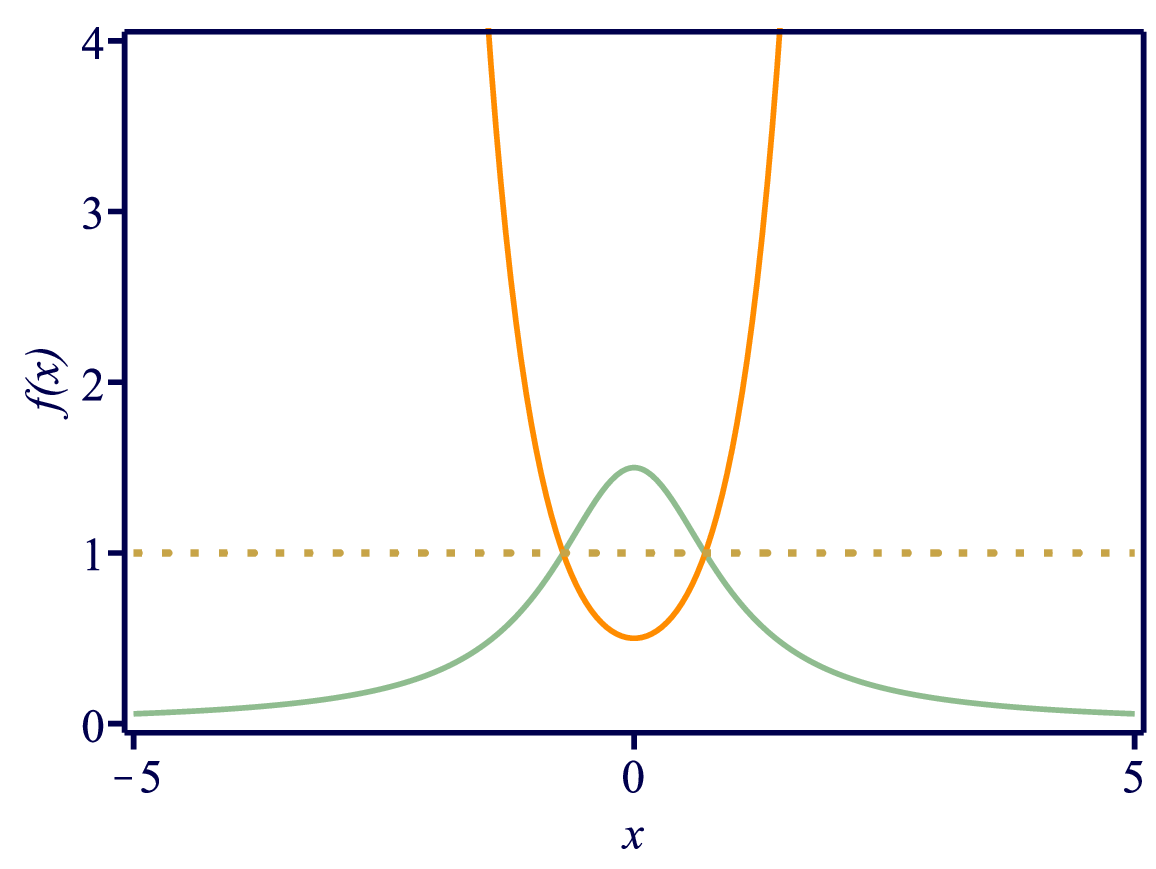}
\includegraphics[width=7.5cm,clip]{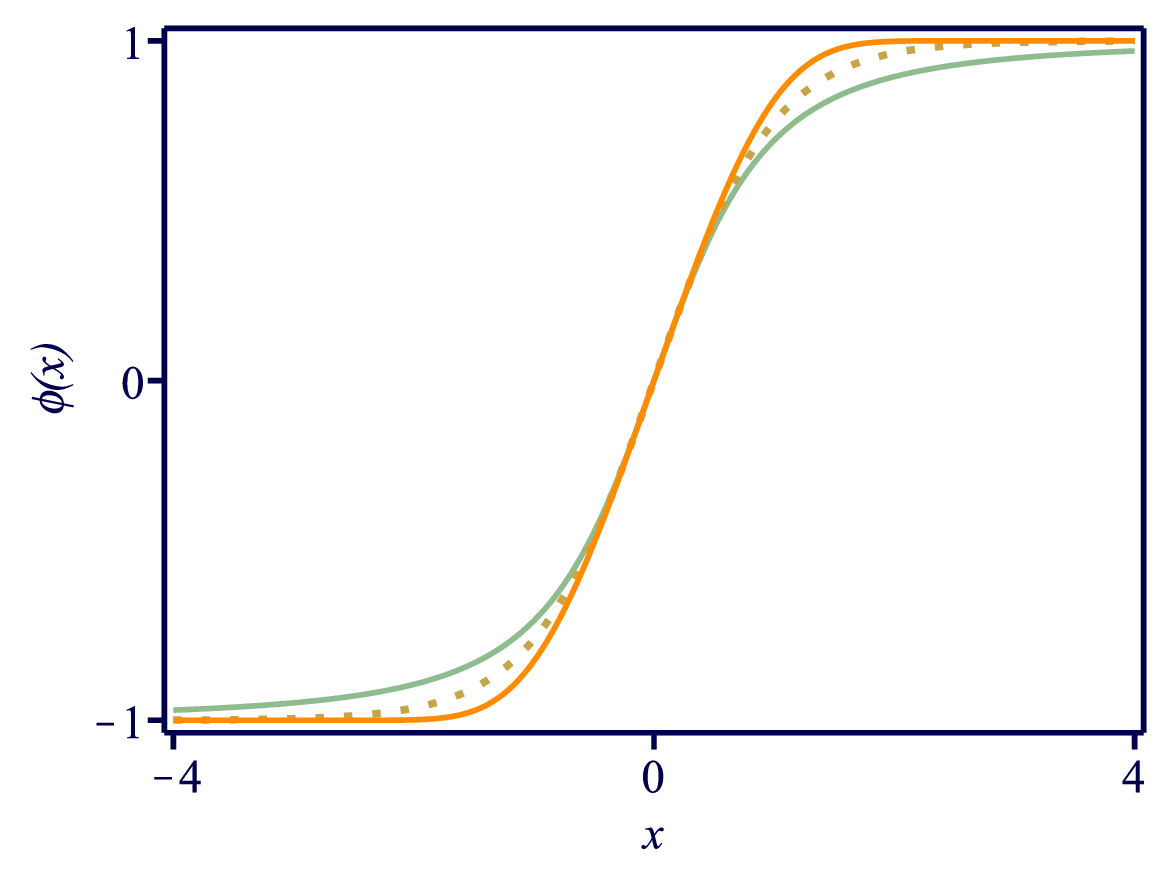}
\includegraphics[width=7.5cm,clip]{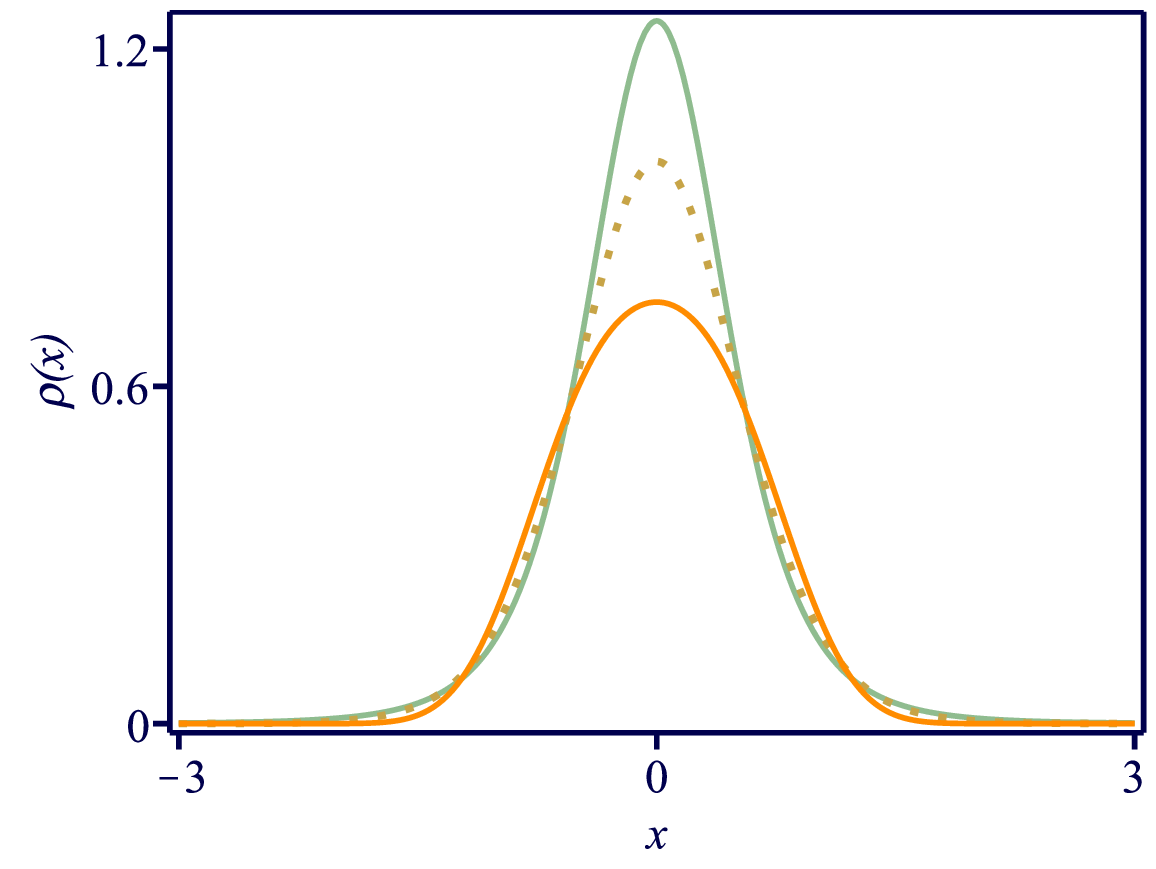}
\includegraphics[width=7.5cm,clip]{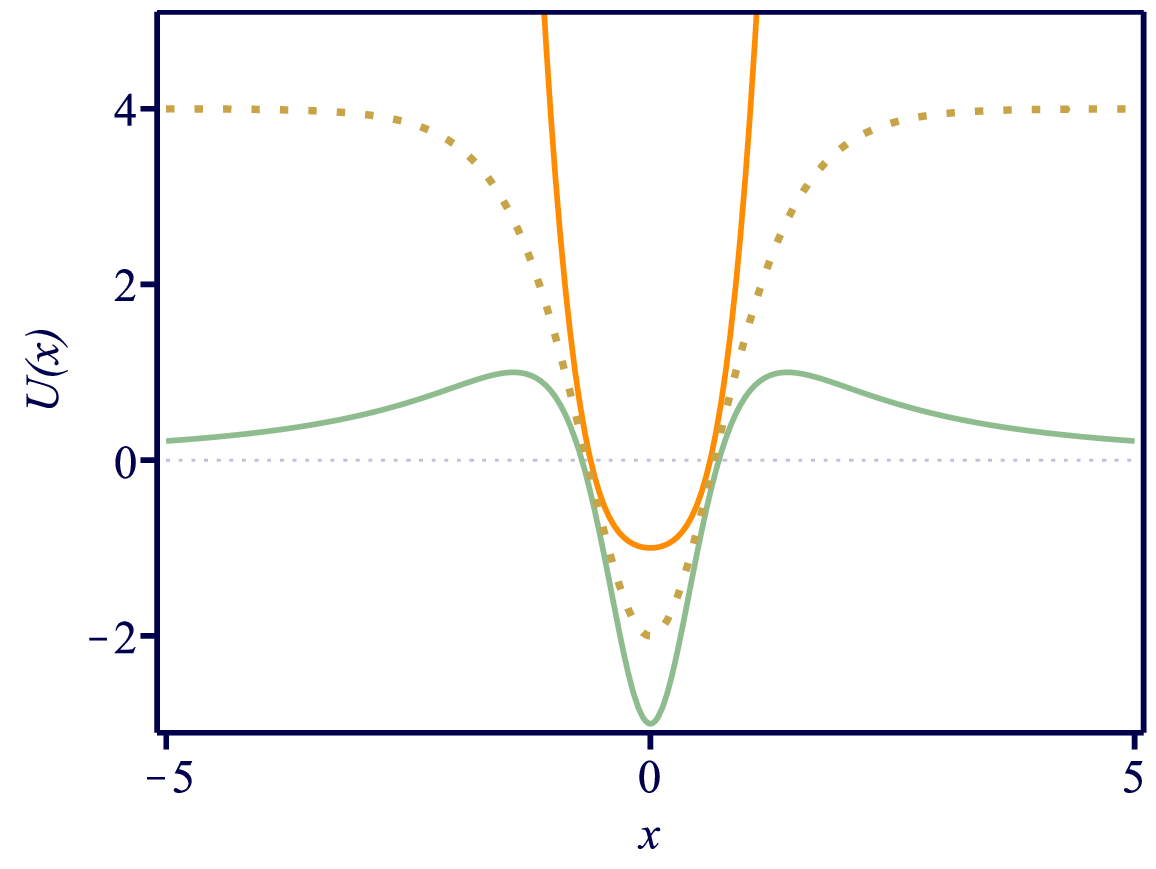}
\caption{The function $f(x)$ in Eqs.~\eqref{impuritylong} and \eqref{impuritydouble} (top left; green and orange, respectively), the solutions $\phi(x)$ in Eqs.~\eqref{sollong} and \eqref{soldouble} (top right; green and orange, respectively), the energy densities $\rho(x)$ in Eqs.~\eqref{rholong} and \eqref{rhodouble} (bottom left; green and orange, respectively), and the stability potentials $U(x)$ in Eqs.~\eqref{ulong} and \eqref{udouble} (bottom  right; green and orange, respectively). The dotted lines represent the aforementioned quantities for the impurity-free case, in which ${\cal I}(x)=0$ or $f(x)=1$, and are used as reference in each panel.}
\label{figlongdouble}
\end{figure}

We start the investigation of the stability of the solution \eqref{sollong} with the Derrick's argument and its extension, which lead to the conditions \eqref{conditionimp12} and \eqref{derricktranslim}. Although the system is stable against contractions and dilations, obeying Eq.~\eqref{conditionimp12}, it is unstable against translations, violating the conditions \eqref{derricktranslim}. The instability of the system also arises from the investigation of the linear stability, described by Eq.~\eqref{stabilityim}, whose associated stability potential \eqref{stabpot} reads
\be\label{ulong}
U(x) = \frac{6x^2-3}{\left(x^2+1\right)^2}.
\ee
It has a volcano profile, such that, $U(\pm\infty)=0$. By analyzing its associated stability equation \eqref{stabilityim}, we have found the presence of the negative eigenvalue $\omega^2\approx-0.6596$. Thus, although the solution \eqref{sollong} has a kink profile, the presence of the impurity driven by the function in Eq.~\eqref{impuritylong} leads to instability. In Fig.~\ref{figlongdouble}, we display function \eqref{impuritylong}, the solution \eqref{sollong}, the energy density \eqref{rholong} and the stability potential \eqref{ulong}. The expression for the charge density \eqref{chargedens} is cumbersome, so we omit it here. From Eq.~\eqref{eeps}, we get the electric field
\be
\textbf{E} = \frac{\sqrt{6}\,\text{sgn}(x)}{2\left(x^2+1\right)^{5/2}}\hat{x}.
\ee
Notice that this electric field has power law tails, as expected.

The scalar field solution with power law tails can be used in collisions, to see how the long-range interaction behaves in the present case. This is of current interest, and may help us better understand the physics involved in Rydberg atom and in cold atoms in cavities; see, e.g., Ref. \cite{Cold} and references therein.

\subsubsection{Quasi-compact structures}

Previously, we discussed conditions for $f(x)$ under which the scalar field has double exponential tails. Objects with this feature have been investigated in the literature and, due to their rapid falloff, were called quasi-compact structures \cite{quasi1,quasi2,quasi3}. To illustrate this situation, we take
\be\label{impuritydouble}
f(x) = \cosh^2(x)-\frac{1}{2}\sinh(x)\coth\left(\sinh(x)\right).
\ee
This function diverges at $x\to\pm\infty$ and has a minimum at the origin, such that $f(0)=1/2$. It somehow simulates a very deep well, which will be able to entrap the solution around the origin. The static solution of Eq.~\eqref{phiV} connecting the minima of the potential \eqref{phi4} is
\be\label{soldouble}
\phi(x) = \tanh\left(\sinh(x)\right).
\ee
Since $f(x)\approx \exp(2x)/4$ asymptotically, for very large values of $x$, we see that this solution has double exponential tails, as expected. The energy density \eqref{rhoim} reads
\be\label{rhodouble}
\rho(x) = \sech^4\left(\sinh(x)\right)\left(\cosh^2(x)-\frac{1}{4}\sinh(x)\coth\left(\sinh(x)\right)\right).
\ee
By integrating the above expression, one gets the energy ${\cal E} =1.1888$. The stress from Eq.~\eqref{sigmaim} becomes $\sigma(x) = \sinh(x)\coth\left(\sinh(x)\right)\sech^4\left(\sinh(x)\right)/4$.

To investigate the stability, we first consider the conditions \eqref{conditionimp12} and \eqref{derricktranslim}. By using a numerical approach, we have checked that they are obeyed, so the system is stable under rescale and spatial translations. We also study the linear stability using Eq.~\eqref{stabilityim}, with  Eq.~\eqref{stabpot} leading to the stability potential 
\be\label{udouble}
U(x) = \left(4 -6\,\sech^2\left(\sinh(x)\right)\right)\left(\cosh^2(x) -\frac{1}{2}\sinh(x)\coth\left(\sinh(x)\right)\right).
\ee
We then use numerical simulation to conclude that there are infinite eigenvalues associated to the above potential and all of them are positive, $\omega_i^2>0$. In Fig.~\ref{figlongdouble}, we display the function \eqref{impuritydouble}, the solution \eqref{soldouble}, the energy density \eqref{rhodouble} and the stability potential \eqref{udouble}. We can compare these quantities with the corresponding ones which appear in the models investigated in Secs.~\ref{secpoint} and \ref{seclongrange}.

The charge configuration associated to the impurity governed by the function \eqref{impuritydouble} is omitted, as in the previous model, because it is cumbersome. We remark, however, that $\delta(x)$ appears in $\varrho(x)$ since $f(0)\neq0$. In this case, the electric field \eqref{eeps} becomes
\be
\textbf{E} = \sech^4(\sinh(x))\,\sqrt{\cosh^2(x)-\frac{1}{2}\sinh(x)\coth\left(\sinh(x)\right)}\,\text{sgn}(x)\,\hat{x}.
\ee
This electric field is localized, behaving asymptotically as $\textbf{E} = 8\exp(-2\exp(x))\,\hat{x}$ for very large values of $x$, highlighting the quasi-compact profile of the structure.

\section{Conclusions}\label{secconclusions}
In this work, we have investigated the Maxwell-scalar system \eqref{lel} in $(1,1)$ dimensions with electric permittivity governed by a real scalar field. We have shown that the equation of motion that drives the scalar field is independent from the gauge field and have calculated the physical quantities associated to the system, such as the energy, momentum, stress and electric field. Furthermore, we have studied the stability of static configurations around small fluctuations and under rescale and spatial translations. We have also presented the model \eqref{lim}, which contains a single real scalar field in the presence of impurities. Interestingly, if the conditions \eqref{condeff} are satisfied, the model \eqref{lim} can be seen as an effective model for the Maxwell-scalar system \eqref{lel}. The results suggest an interesting interpretation for the impurity, since it can be used to describe charge density in the Maxwell-scalar system. Thus, we can consider the model \eqref{lim} to get the scalar field and then calculate the electric field associated to \eqref{lel} for the electrically charged model.

This correspondence between the Maxwell-scalar system and the scalar field model with impurity is the major result of this work, and we have illustrated it within several distinct scenarios, to find nontrivial solutions that confirm the robustness of the result. We also emphasize here that the unveiled correspondence is obtained analytically, without relying on numerical simulations. In this sense, this work provides exact one-dimensional solutions that can perhaps be further considered under the emerging area of quantum simulation of the fundamental quantum fields in high energy physics \cite{QC,QC2,QS}.

As a direct perspective, the model \eqref{lim} can be investigated with the presence of more general impurities, in which asymmetries may play specific role, requiring the use of numerical procedures. We can also study the scalar field model in a periodic lattice of impurities, in a way similar to the discussions considered before in Ref. \cite{imp3}. Another line of current interest concerns the
use of a complex scalar field in the scalar field model and also, in the Maxwell-scalar system, to change the external charge density for a dynamical complex scalar field to
be minimally coupled with the gauge field. 
This may induce the presence of gauged Q-balls; see, e.g, Refs. \cite{QB,QB2}.
Another interesting scenario is related to the case of a larger number of scalar fields, in a multi-field environment, with the extra degrees of freedom used to control the internal structure of the solutions. The multi-field extension can be implemented in the Maxwell-scalar system and in the scalar field model, hopefully enriching the results described in this work. It is also possible to explore the correspondence using higher spatial dimensions. In this direction, we can view distinct possibilities, one of them being the search of models with impurities to verify if the equivalence still holds in the electrostatic scenario. A concrete line of study could be related to the work \cite{newglob}, in which the authors constructed stable kink-like configurations in higher spatial dimensions. Another possibility would consider investigating models in curved spacetime, for instance, around black holes and other background geometries, as in Refs.~\cite{morris1,morris2,danilo,danilo2}. We can also consider the extension including complex scalar fields in $(2,1)$ spacetime dimensions, which may be useful to study vortices in the presence of magnetic impurities \cite{vortimp,vortimp2,vortimp3}. Some of these issues are under consideration, and we hope to report on them in the near future.

 \acknowledgements{This work is supported by the Brazilian agency Conselho Nacional de Desenvolvimento Cient\'ifico e Tecnol\'ogico (CNPq), grants 303469/2019-6 (DB), 306151/2022-7 (MAM) and 310994/2021-7 (RM), and by Paraiba State Research Foundation (FAPESQ-PB) grants 0003/2019 (RM) and 0015/2019 (DB and MAM).}


\begin{thebibliography}{99}
\bibitem{jackson}J.D. Jackson, \textit{Classical Electrodynamics}, Wiley (1999).
\bibitem{landau}L.D. Landau and E.M. Lifshitz, \textit{The Classical Theory of Fields}, Butterworth-Heinemann (1980).
\bibitem{S1}J. Schwinger, \textit{Gauge Invariance and Mass}, Phys. Rev. {\bf125} (1962) 397.
\bibitem{S2}J. Schwinger, \textit{Gauge Invariance and Mass. II}, Phys. Rev. {\bf128} (1962) 2425.
\bibitem{QC}N. Klco, E.F. Dumitrescu, A.J. McCaskey, T.D. Morris, R.C. Pooser, M. Sanz, E. Solano, P. Lougovski, M.J. Savage, \textit{Quantum-classical computation of Schwinger model dynamics using quantum computers.} Phys. Rev. A 98 (2018) 032331.


\bibitem{eletricloc} D. Bazeia, M.A. Marques and R. Menezes, \textit{Electrically charged localized structures}, Eur. Phys. J. C {\bf81} (2021) 94.
\bibitem{dipole}D.~Bazeia, M.A.~Marques and R.~Menezes, \textit{Maxwell-scalar device based on the electric dipole}, Phys. Rev. D \textbf{104} (2021) L121703. 
\bibitem{morris2}J.R.~Morris, \textit{BPS equations and solutions for Maxwell\textendash{}scalar theory}, Annals Phys. \textbf{438} (2022) 168782.	

\bibitem{manton}N. Manton and P. Sutcliffe, \textit{Topological Solitons}, Cambridge University Press (2004).
\bibitem{vachaspati}T. Vachaspati, \textit{Kinks and Domain Walls: An Introduction to Classical and Quantum Solitons}, Cambridge University Press (2006).
\bibitem{bogo}E.B. Bogomol'nyi, \textit{Stability of Classical Solutions}, Sov. J. Nucl. Phys. \textbf{24} (1976) 449.
\bibitem{PS}M.K. Prasad and C.M. Sommerfield, {\textit{ Exact Classical Solution for the 't Hooft Monopole and the Julia-Zee Dyon}}, Phys. Rev. Lett. {\bf35} (1975) 760.
\bibitem{rs}L. Randall and R. Sundrum, \textit{An Alternative to Compactification}, Phys. Rev. Lett. {\bf83} (1999) 4690.
\bibitem{dewolfe}O. DeWolfe, D.Z. Freedman, S.S. Gubser and A. Karch, \textit{Modeling the fifth dimension with scalars and gravity}, Phys. Rev. D {\bf62} (2000) 046008.
\bibitem{csaki} C. Csaki, J. Erlich, T.J. Hollowood and Y. Shirman, \textit{Universal aspects of gravity localized on thick branes}, Nucl. Phys. B {\bf581} (2000) 309.
\bibitem{zhong} Y. Zhong, \textit{Revisit on two-dimensional self-gravitating kinks: superpotential formalism and linear stability}, J. High Energ. Phys. {\bf2021} (2021) 118.
\bibitem{bec1}J. Belmonte-Beitia, V.M. Perez-Garcia, V. Vekslerchik, V.V. Konotop, \textit{Localized Nonlinear Waves in Systems with Time- and Space-Modulated Nonlinearities}, Phys. Rev. Lett. {\bf100} (2008) 164102 .
\bibitem{bec2}A.T. Avelar, D. Bazeia, W.B. Cardoso, \textit{Solitons with Cubic and Quintic Nonlinearities Modulated in Space and Time}, Phys. Rev. E {\bf79} (2009) 025602(R).
\bibitem{mag1}P.-O. Jubert, R. Allenspach and A. Bischof, \textit{Magnetic domain walls in constrained geometries}, Phys. Rev. B {\bf69} (2004) 220410(R).
\bibitem{mag2} A. Vanhaverbeke, A. Bischof and R. Allenspach, \textit{Control of Domain Wall Polarity by Current Pulses}, Phys. Rev. Lett. {\bf101} (2008) 107202.

\bibitem{imp1}J.F. Currie, S.E. Trullinger, A. R. Bishop, and J. A. Krumhansl, \textit{Numerical simulation of sine-Gordon soliton dynamics in the presence of perturbations}, Phys. Rev. B {\bf15} (1977) 5567.
\bibitem{imp2}D.W. McLaughlin and A. C. Scott, \textit{Perturbation analysis of fluxon dynamics}, Phys. Rev. A {\bf18} (1978) 1652.
\bibitem{imp3}Y.S. Kivshar and B.A. Malomed, \textit{Dynamics of solitons in nearly integrable systems}, Rev. Mod. Phys. {\bf61} (1989) 763.
\bibitem{imp4}Y.S. Kivshar, Z. Fei and L. V\'azquez, \textit{Resonant soliton-impurity interactions}, Phys. Rev. Lett. {\bf67} (1991) 1177.
\bibitem{imp5} Z. Fei, Y.S. Kivshar and L. V\'azquez, \textit{Resonant kink-impurity interactions in the sine-Gordon model}, Phys. Rev. A {\bf45} (1992) 6019.
\bibitem{imp6} Z. Fei, Y.S. Kivshar and L. V\'azquez, \textit{Resonant kink-impurity interactions in the $\varphi^4$ model}. Phys. Rev. A {\bf46} (1992) 5214.
\bibitem{imp7} Z. Fei, V.V. Konotop, M. Peyrard and L. Vázquez, \textit{Kink dynamics in the periodically modulated $\varphi^4$ model}, Phys. Rev. E {\bf48} (1993) 548.
\bibitem{imp8} G. K\"albermann, \textit{Soliton tunneling}, Phys. Rev. E {\bf55} (1997) R6360(R).
\bibitem{imp9} J. Garnier and F.Kh. Abdullaev, \textit{Transmission of matter-wave solitons through nonlinear traps and barriers}, Phys. Rev. A {\bf74} (2006) 013604.
\bibitem{imp10}C. Adam and A. Wereszczynski, \textit{BPS property and its breaking in $1+1$ dimensions}, Phys. Rev. D {\bf98} (2018) 116001.
\bibitem{imp11} C. Adam, T. Romanczukiewicz, and A. Wereszczynski, \textit{The $\phi^4$ model with the BPS preserving defect}, J. High Energ. Phys. {\bf2019} (2019) 131. 
\bibitem{imp12} C. Adam, K. Oles, J. Queiruga T. Romanczukiewicz, A. Wereszczynski, \textit{Solvable self-dual impurity models}, J. High Energ. Phys. {\bf2019} (2019) 150.
\bibitem{imp13}C. Adam, J.M. Queiruga, A. Wereszczynski, \textit{BPS soliton-impurity models and supersymmetry}, J. High Energ. Phys. {\bf2019} (2019) 164. 
\bibitem{imp14}J.G.F. Campos and A. Mohammadi, \textit{Fermion transfer in the $\phi^4$ model with a half-BPS preserving impurity}, Phys. Rev. D \textbf{102} (2020) 045003.

\bibitem{imp15}I. Takyi and H. Weigel, \textit{Quantum effects of solitons in the self-dual impurity model}, Phys. Rev. D \textbf{107} (2023) 036003.

\bibitem{PRL} C. Adam, K. Oles, T. Romanczukiewicz, A. Wereszczynski, \textit{Spectral Walls in Soliton Collisions}, Phys. Rev. Lett. {\bf122} (2019) 241601.

\bibitem{PRD}C. Adam, K. Oles, T. Romanczukiewicz, A. Wereszczynski, \textit{Kink-antikink scattering in the $\varphi^4$ model without static intersoliton forces}, Phys. Rev. D {\bf101} (2020) 105021.

\bibitem{SW1}C. Adam, K. Oles, T. Romanczukiewicz, A. Wereszczynski, and W. J. Zakrzewski, \textit{Spectral walls in multifield kink dynamics}, J. High Energ. Phys. {\bf2021} (2021) 147.



\bibitem{SW2}João G.F. Campos, Azadeh Mohammadi, Jose M. Queiruga, Andrzej Wereszczynski, W.J. Zakrzewski, \textit{Fermionic spectral walls in kink collisions}, J. High Energ. Phys. {\bf2023} (2023) 071.


\bibitem{QC2}A.C.Y. Li, A. Macridin, S. Mrenna, P. Spentzouris, \textit{Simulating scalar field theories on quantum computers with limited resources.} Phys. Rev. A 107 (2023) 032603.

\bibitem{ads1}J. M. Maldacena, \textit{The Large N Limit of Superconformal field theories and supergravity}, Adv. Theor. Math. Phys. 2 (1998) 231.

\bibitem{ads2}J. Maldacena, \textit{
The Large N limit of superconformal field theories and supergravity}, Int. J. Theor. Phys. 38 (1999) 1113.

 
\bibitem{ads3}S. S. Gubser, I. R. Klebanov, A. M. Polyakov, \textit{Gauge Theory Correlators from Non-Critical String Theory}, Phys. Lett. B 428 (1998) 105.

\bibitem{ads4}E. Witten, \textit{Anti De Sitter Space And Holography}, Adv. Theor. Math. Phys. 2 (1998) 253.

\bibitem{jac1}D. Bazeia and R. Jackiw, \textit{Nonlinear realization of a dynamical Poincare symmetry by a field-dependent diffeomorphism}, Ann. Phys. (NY) 270 (1998) 246.

\bibitem{B1}D. Bazeia, \textit{Galileo invariant system and the motion of relativistic d-branes}, Phys. Rev. D 59 (1999) 085007.

\bibitem{jac2}R. Jackiw and A.P. Polychronakos, \textit{Fluid Dynamical Profiles and Constants of Motion from d-Branes}, Comm. Math. Phys. 207 (1999) 107.


\bibitem{jac}R. Jackiw, \textit{Lectures on Fluid Dynamics: A Particle Theorist’s View of Supersymmetric, Non-Abelian, Noncommutative Fluid Mechanics and d-Branes}, Springer (2002).

\bibitem{olmo1}V. I. Afonso, G. J. Olmo, E. Orazi, D. Rubiera-Garcia, \textit{Correspondence between modified gravity and general relativity with scalar fields},  Phys. Rev. D 99 (2019) 044048.

\bibitem{olmo2}A. Masó-Ferrando, N. Sanchis-Gual, J. A. Font, G. J. Olmo, \textit{Birth of baby universes from gravitational collapse in a modified-gravity scenario,} J. Cosmology Astropart. Phys. 06 (2023) 028.

\bibitem{Cold}N. Defenu, T. Donner, T. Macrì, G. Pagano, S. Ruffo, A. Trombettoni, \textit{Long-range interacting quantum systems,}
Rev. Mod. Phys. 95 (2023) 035002.

\bibitem{derrick} G.H. Derrick, \textit{Comments on Nonlinear Wave Equations as Models for Elementary Particles}, J. Math. Phys. \textbf{5} (1964) 1252.

\bibitem{QBB}D. Bazeia, M.A. Marques, R. Menezes, \textit{Exact solutions, energy, and charge of stable Q-balls}, Eur. Phys. J. C (76 2016) 241.

\bibitem{QB}K. Lee, J.A. Stein-Schabes, R. Watkins, L.M. Widrow, \textit{Gauged Q-balls}, 
Phys. Rev. D 39 (1989) 1665.

\bibitem{QB2}A.Yu. Loginov, V.V. Gauzshtein, \textit{One-dimensional soliton system of gauged kink and Q-ball}, Eur. Phys. J. C 79 (2019) 780. 

\bibitem{DB}D. Bazeia, \textit{Defect Structures in Field Theory}, [arXiv:hep-th/0507188].

\bibitem{stabkink}I.~Andrade, M.A.~Marques and R.~Menezes, \textit{Stability of kinklike structures in generalized models}, Nucl. Phys. B \textbf{951} (2020) 114883.


\bibitem{newglob} D. Bazeia, J. Menezes and R. Menezes, \textit{New Global Defect Structures},
	Phys. Rev. Lett. {\bf91} (2003) 241601.

\bibitem{highly}A.R.~Gomes, R.~Menezes and J.C.R.E.~Oliveira, \textit{Highly interactive kink solutions}, Phys. Rev. D \textbf{86} (2012) 025008.


\bibitem{quasi1}G.~Flores-Hidalgo and N.F.~Svaiter, \textit{Constructing bidimensional scalar field theory models from zero mode fluctuations}, Phys. Rev. D \textbf{66} (2002) 025031.

\bibitem{quasi2}D.~Bazeia, M.A.~Marques and R.~Menezes, \textit{Quasi-compact Q-balls}, EPL \textbf{127} (2019) 21001.
\bibitem{quasi3}
D.~Bazeia, M.~A.~Marques and R.~Menezes, \textit{Quasi-compact vortices}, EPL \textbf{129} (2020) 31001.

\bibitem{QS}C.W. Bauer, Z. Davoudi, N. Klco, M.J. Savage, \textit{Quantum simulation of fundamental particles and forces,}  Nature Rev. Phys. 5 (2023) 420.


\bibitem{morris1}J.R.~Morris, \textit{Radially symmetric scalar solitons}, Phys. Rev. D \textbf{104} (2021) 016013.

\bibitem{danilo2}D.C.~Moreira, F.A.~Brito and D.~Bazeia, \textit{Localized scalar structures around static black holes}, Nucl. Phys. B \textbf{987} (2023) 116090.

\bibitem{danilo}D.C.~Moreira, F.A.~Brito and J.C. Mota-Silva, \textit{Scalar fields and Lifshitz black holes from Derrick’s theorem evasion}, Phys. Rev. D \textbf{106} 106 (2022) 125017.

\bibitem{vortimp}D. Tong and K. Wong, \textit{Vortices and Impurities}, J. High Energ. Phys. {\bf2014} (2014) 90.

\bibitem{vortimp2}J. Ashcroft and S. Krusch, \textit{Vortices and magnetic impurities}, Phys. Rev. D {\bf101} (2020) 025004.

\bibitem{vortimp3}D.~Bazeia, M.A.~Liao and M.A.~Marques, \textit{Impurity-like solutions in vortex systems coupled to a neutral field}, Phys. Lett. B \textbf{825} (2022) 136862.



\end{thebibliography}
\end{document}